\documentclass[usenatbib,usegraphicx]{mn2e}
\newcommand{\be}{\begin{equation}}
\newcommand{\ee}{\end{equation}}
\newcommand{\bea}{\begin{eqnarray}}
\newcommand{\eea}{\end{eqnarray}}
\newcommand{\bc}{\begin{center}}
\newcommand{\ec}{\end{center}}

\def\gsim{ \lower .75ex \hbox{$\sim$} \llap{\raise .27ex \hbox{$>$}} }
\def\lsim{ \lower .75ex \hbox{$\sim$} \llap{\raise .27ex \hbox{$<$}} }

\renewcommand{\thefootnote}{\fnsymbol{footnote}}

\title[Gravitational recoils of supermassive black holes in gas rich galaxies]{Gravitational recoils of supermassive black holes in hydrodynamical
  simulations of gas rich galaxies}

\author[Sijacki et al.]
       {\parbox{18cm}{Debora~Sijacki$^{1,2}$\footnotemark[1], Volker
           Springel$^{3,4}$ and Martin~G. Haehnelt$^{1}$}\vspace{0.3cm}\\ 
         $^1$ Kavli Institute for Cosmology, Cambridge and Institute of
         Astronomy, Madingley Road, Cambridge, CB3 0HA \\
         $^2$ Harvard-Smithsonian Center for Astrophysics, 60 Garden Street,
         Cambridge, MA, 02138, USA\\
         $^3$ Heidelberg Institute for Theoretical Studies, Schloss-Wolfsbrunnenweg 35, 69118 Heidelberg, Germany\\
         $^4$ Zentrum f\"{u}r Astronomie der Universit\"{a}t Heidelberg, ARI, M\"onchhofstr. 12-14, 69120 Heidelberg, Germany}

\begin{document}

\maketitle
\begin{abstract} 

We study the evolution of gravitationally recoiled supermassive black holes
(BHs) in massive gas-rich galaxies by means of high-resolution hydrodynamical
simulations. We find that the presence of a massive gaseous disc allows
recoiled BHs to return to the centre on a much shorter timescale than for
purely stellar discs. Also, BH accretion and feedback can strongly modify the
orbit of recoiled BHs and hence their return timescale, besides affecting the
distribution of gas and stars in the galactic centre. However, the dynamical
interaction of kicked BHs with the surrounding medium is in general complex
and can facilitate both a fast return to the centre as well as a significant
delay. The Bondi-Hoyle-Lyttleton accretion rates of the recoiling BHs in our
simulated galaxies are favourably high for the detection of off-centred AGN if
kicked into gas-rich discs -- up to a few per cent of the Eddington accretion
rate -- and are highly variable on timescales of a few $10^7$ yrs.  In major
merger simulations of gas-rich galaxies, we find that gravitational recoils
increase the scatter in the BH mass -- host galaxy relationships compared to
simulations without kicks, with the BH mass being more sensitive to recoil
kicks than the bulge mass. The BH mass can be lowered by a factor of a few due
to a recoil, even for a relatively short return timescale, but the exact
magnitude of the effect depends strongly on the BH binary hardening timescale
and on the efficiency of star formation in the central regions. A generic
result of our numerical models is that the clumpy massive discs suggested by
recent high-redshift observations, as well as the remnants of gas-rich
mergers, exhibit a gravitational potential that falls steeply in the central
regions, due to the dissipative concentration of baryons. As a result,
supermassive BHs should only rarely be able to escape from massive galaxies at
high redshifts, which is the epoch where the bulk of BH recoils is expected to
occur.

\end{abstract}


\section{Introduction}

\renewcommand{\thefootnote}{\fnsymbol{footnote}}
\footnotetext[1]{Hubble Fellow. E-mail: dsijacki@cfa.harvard.edu}

In the last couple of years, numerical relativity simulations of coalescing
black hole (BH) binaries have opened a whole new window in our understanding
of the BH properties in the merger aftermath \citep[e.g.][]{Pretorius2007,
  Gonzalez2007, Herrmann2007, Koppitz2007, Campanelli2007, Baker2008}. These
computations have established that for asymmetries in the mass and/or spin of
the two merging BHs, net linear momentum will be carried away by the
asymmetric emission of gravitational waves, imparting a gravitational recoil
to the remnant BH. In the case of the spinning BHs, the recoil velocities can
become very large -- comparable to or even larger than the virial velocities
of the most massive host systems. This is what makes this phenomenon
astrophysically very interesting and relevant, and poses important questions
to which we still do not have definite answers: How often do the recoils occur
within the hierarchical structure formation scenario?  What is the fraction of
expelled BHs, and how fast can they sink back to the centre? What are the
imprints of gravitationally recoiled BHs and why has it been difficult to
detect them so far?

The importance of BH recoils in the astrophysical context has been realised in
pioneering works early on \citep[e.g.][]{Bekenstein1973, Blandford1979,
  Kapoor1985}.  More recently, analytical and numerical studies
\citep[e.g.][]{Merritt2004, Boylan2004, Madau2004, Micic2006} have found that globular
clusters and dwarf galaxies are the prime targets for getting depleted of
their central BHs. Moreover, recoiled BHs which do not escape their host
galaxies are likely to induce stellar cores through repeated passages close to
the centre.

In the case of purely stellar systems, accurate N-body simulations \citep[for
  a recent study see][]{Gualandris2008} have led to a good understanding how
gravitationally recoiled BHs orbit in spherical systems and what their typical
return timescales for different kick velocities are. For systems with gas,
however, only a handful of numerical studies \citep[e.g.][]{Kornreich2008,
  Devecchi2009, Guedes2011} are available. These studies computed the trajectories of
recoiled BHs and accounted for their possible interaction with the surrounding
gas. Adopting a semi-analytical approach, BH luminosities during the wandering
phase have nevertheless been estimated \citep[e.g.][]{VolonteriPerna2005,
  Blecha2008, Fujita2009, Guedes2011}. Recently, \citet{Dotti2010}
  performed simulations of BH binaries embedded in a circum-nuclear disc
  allowing for gas accretion, and by analytically tracing BH
  spin evolution, found that high recoil velocities should not be very likely.
While observationally several  candidates for recoiled BHs have been proposed
(\citet{Komossa2008, Civano2010}, but see \citet{Shields2009, Heckman2009}),
there is scarce evidence for luminous recoiled quasars, at least in the case
of large kinematic offsets \citep{Bonning2007}.

For galaxies containing gas, the evolution of recoiled BHs can possibly vary
significantly depending on the mass fraction of the gas, and its spatial
distribution and thermodynamic state. This situation presents a much more
complex problem than for purely stellar systems.  For example, recoiled BHs
might accrete some of the interstellar gas which could in return affect their
trajectory. Also, if BH feedback effects associated with accretion are not
negligible, they could provide distributed heating throughout the galaxy. It
is clear that a full understanding of these possibilities will require the
exploration of a large parameter space.

At redshifts of $z \sim 2-3$, galaxy mergers are much more common in the
hierarchical structure formation scenario than at the present day, and it is
believed that many galaxies are gas-rich, with gas fractions of up to $50\%$
\citep[][and references therein]{Forster2009}. Massive compact ellipticals at
$z \sim 2$ with much smaller sizes and higher stellar densities than their
local counterparts \citep[e.g.][]{Dokkum2008} are likely to be the end products
of gas-rich mergers where remnant BHs would reside. This epoch also
corresponds to the peak of the quasar space density, indicating that BHs are
accreting gas supplied by their hosts efficiently. This is hence an extremely
important and interesting regime, where BH mergers and recoils are expected to
occur frequently and at the same time the gas component in the host systems
cannot be neglected.

In this paper, we present an exploratory study of the properties of recoiled
BHs in gas-rich galaxies, where BH accretion and feedback processes are
followed self-consistently. We first focus on studying an isolated spiral
galaxy, simulated at a high resolution to understand how the inclusion of gas
accretion or additional BH feedback affects the orbits of recoiled BHs, their
return timescale to the centre, and their imprints on the host galaxy. We
simulate kicks of different magnitude, both in the plane and perpendicular to
the plane of the galaxy, to gauge the dependence of the AGN luminosity on the
kick orientation and on the assumed equation of state for the interstellar
gas. We then perform a major merger simulation of two gas-rich galaxies, each
containing a supermassive BH in its centre. After BH coalescence, we follow
the evolution of the gravitationally recoiled BH and study how its growth is
affected by the kick, and which consequences this has for the scatter in the
BH mass-host galaxy scaling relations. Note that a systematic study of recoiled BHs in
merging galaxies is underway \citep[][private communication]{Blecha2011}.

The paper is organised as follows. In Section~\ref{Methodology}, we outline
the numerical methods we adopted. Most of our results are presented in
Section~\ref{Results}, where we discuss isolated spiral galaxies with uniform
or clumpy discs (Section~\ref{isolated}), and major mergers of two gas-rich
galaxies (Section~\ref{merging}). In Section~\ref{Conclusions}, we discuss our
findings and draw our conclusions.

\section{Methodology} \label{Methodology}

\begin{figure}\centerline{
\includegraphics[width=8.3truecm,height=8.truecm]{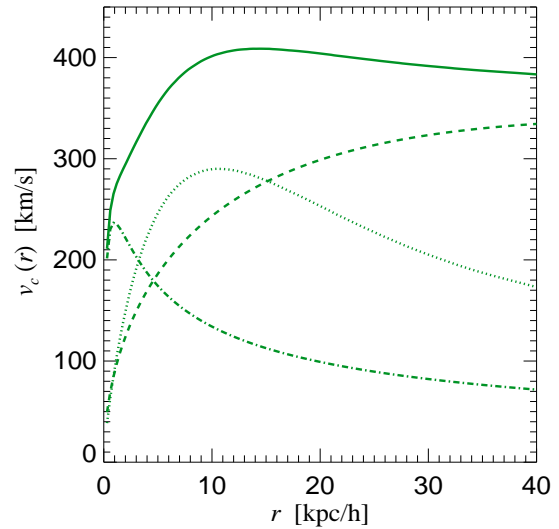}}
\vspace{-0.5cm}
\caption{Circular velocity curves of the isolated galaxy with $v_{\rm 200}
  = 300\, {\rm km\, s^{-1}}$ for different components: bulge (dot-dashed
  line), disc (dotted line), dark matter halo (dashed line), and total
  (continuous line).}  
\label{vcirc}
\end{figure}

\begin{figure}\centerline{
\includegraphics[width=8.truecm,height=8.truecm]{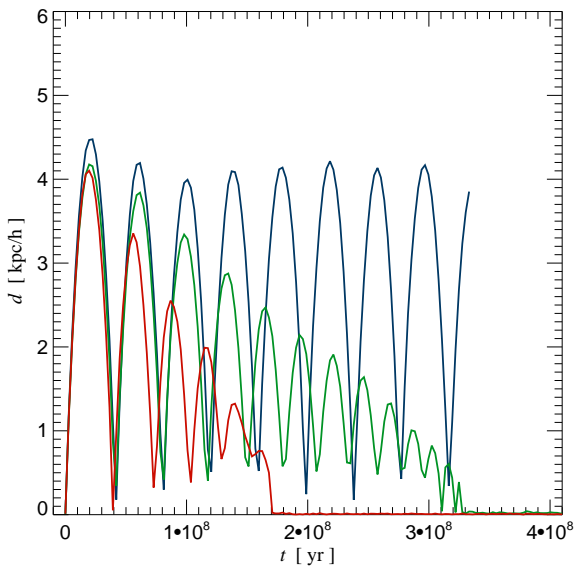}}
\caption{BH distance from the minimum of the potential as a function of time
  for simulations where the BH is (a) not subject to dynamical friction
  forces (blue line), (b) experiences dynamical friction from stars only (green
  line), and (c) additionally is subject to dynamical friction from the gas in the
  disc (red line).}  
\label{BHdist_drag}
\end{figure}

\subsection{The numerical code}

In this work we perform hydrodynamical simulations with the massively parallel
Tree-SPH code {\small GADGET-3} \citep[last described in][]{Gadget2}. The code
computes gravitational forces acting on dark matter, gas, star and BH
particles, as well as the hydrodynamical forces that affect the baryons. Gas
is modelled as an optically thin plasma of hydrogen and helium, which can
radiatively cool and heat. Star formation and supernova feedback is
implemented adopting a subresolution multi-phase model \citep{SpringelH2003},
while BH growth and feedback is modelled as described in \citet{DiMatteo2005,
  Springel2005b, Sijacki2007, Sijacki2009}. For completeness, we briefly
summarise this model below.

\begin{table} 
\bc
\begin{tabular}{rrrrrr} 
\hline
$f_{\rm gas}$ & $q_{\rm EOS}$ & $v_{\rm 200}$ & $v_{\rm esc}(0)$ & $v_{\rm kick}$ & $M_{\rm BH}$\\
     &     &$[{\rm km\, s^{-1}}]$ & $[{\rm km\, s^{-1}}]$ & $[{\rm km\,
      s^{-1}}]$ & $[h^{-1}{\rm M}_\odot]$\\
\hline
\multicolumn{6}{c}{\scriptsize UNIFORM DISC} \\
\hline
$0.0$ & $\--$ & $300$ & $1450\;\,$ & $700\;\,$ & $5\times 10^8$\\
$0.6$ & $0.5$ & $300$ & $1450\;\,$ & {$0$, $375$, $700\dagger$} & $5\times 10^8$\\
$0.6$ & $0.5$ & $300$ & $1450\;\,$ & $500\;\,$ & $5\times 10^7$\\
$0.6$ & $0.5$ & $300$ & $1520\ddagger$ & $700\;\,$ & $5\times 10^8$\\
 \hline 
\multicolumn{6}{c}{\scriptsize CLUMPY DISC} \\
\hline
$0.6$ & $0.05$ & $300$ & $1980\;\,$ & {$0$, $1000$, $1600\;\,$} & $5\times 10^8$\\
\hline
\multicolumn{6}{c}{\scriptsize MAJOR MERGER} \\
\hline
$0.6$ & $0.5$ & $378$ & $3510\;\,$ & {$0$, $3253$, $3535\;\,$} & $5\times 10^7$\\
\hline
\end{tabular}
\caption{Gas fraction, equation of state parameter for the interstellar
  medium, virial velocity, central escape velocity, BH recoil velocity and
  initial BH mass for the isolated spiral galaxy with a smooth and a clumpy
  disc, and for the gas-rich major merger. As a standard choice, the gravitational
  recoil direction is within the galactic plane, and the BH is allowed to
  accrete gas and exert feedback. $\dagger$In this case we also simulate kicks
  perpendicular to the galactic plane, and for the recoil in the galactic
  plane we run extra simulations where: i) the BH is massless, ii) the BH does not
  accrete, iii) the BH accretes, but there is no feedback, iv) BH accretion and
  feedback is allowed, but $\alpha\,=\,1$ in 
  equation~(\ref{Bondi_eq}). $\ddagger$This simulation is analogous to the one
  listed in the second row, but the galaxy has been simulated at twice as high
  spatial resolution, resulting in a somewhat deeper central potential and thus
  higher central escape velocity.}
\label{tab_kicks} 
\ec
\end{table}

\subsection{BH model} \label{Default_model}

In the simulation code, BHs are treated as collisionless sink
particles. Starting out with a certain seed mass, BHs can grow in time
by gas accretion, or by merging with other BHs that happen to be
sufficiently close (within each others smoothing lengths) and that
have small relative velocities (less than the local gas sound
speed). The accretion rate is parametrised in terms of a spherically
symmetric Bondi-Hoyle-Lyttleton type accretion flow \citep{Hoyle1939,
  Bondi1944, Bondi1952}, i.e. 
\be 
\dot M_{\rm BH}\,=\, \frac{4\,\pi \,\alpha\, G^2 M_{\rm BH}^2 \,\rho}{\big(c_s^2 +
v^2\big)^{3/2}}\,, 
\label{Bondi_eq} 
\ee where $\alpha$ is a dimensionless parameter, $\rho$ is the gas density,
$c_s$ the sound speed of the gas, and $v$ is the velocity of the BH relative
to the gas. Here, as a default value we fix $\alpha$ to $100$, as in all of
our previous work, because the multiphase model for star formation gives a
comparatively high mean gas temperature as a result of supernova feedback. A
volume-average of the Bondi-rates for the sub-grid cold and hot phases of the
interstellar medium recovers a value of $\alpha$ close to $100$. We have
however found that our results are relatively insensitive to the adopted value
for the $\alpha$ parameter (see the tests in Appendix~\ref{appen}). The
maximum BH accretion rate is limited to the Eddington rate. BH feedback
associated with the accretion is injected in thermal form, where a small
fraction ($5\%$) of the bolometric luminosity is isotropically coupled to the
surrounding gas.

\subsection{Gravitational recoils of BHs}

BH merger remnants are subject to a gravitational recoil due to the
anisotropic emission of gravitational waves, which carry away net linear
momentum. In \citet{Sijacki2009}, we have implemented a simple method to
represent this effect in our simulations. At the moment of a BH binary merger,
the involved BH masses and spins\footnote{Note that in our simulations BH
  spins change only when BHs merge, and any possible spin-up/down due to the
  gas accretion is neglected \citep{King2006, Kesden2010, Dotti2010}.} are
known in the simulation. By making assumptions about the orientation of the
spins with respect to the orbital angular momentum of the BH binary, it is
then possible to compute the expected gravitational kick velocities accurately
based on fitting formulae that are calibrated against numerical relativity
simulations \citep[for a review see][]{Pretorius2007}. In the case of the
isolated galaxies with a single BH, where the primary goal is to understand
the interaction between a recoiling BH and its host, we do not use these
fitting formulae but consider a wide range of possible kick velocities in the
plane and perpendicular to the plane of the galaxy. In the case of the major
merger discussed in Section~\ref{merging} we again consider a range of
physically interesting kick velocities (comparable to the central escape
velocity), and by assuming that the two BHs are close to maximally spinning we
adopt the relations from \citet{Campanelli2007} to estimate possible kick
magnitudes. In our original model \citep{Sijacki2009}, the BH recoil direction
is assumed to be random with respect to the gas distribution of the host,
given that there is no firm observational evidence for any special alignment
of BH jets with the host galaxy structure. Here however we focus on the case
where the BH is recoiled in the plane of the forming circum-nuclear disc. This
turns out to be a physically more interesting case for studying recoiled BH -
host interactions without compromising the general validity of our findings.

\begin{figure*}\centerline{\hbox{
\includegraphics[width=6.5truecm,height=6.truecm]{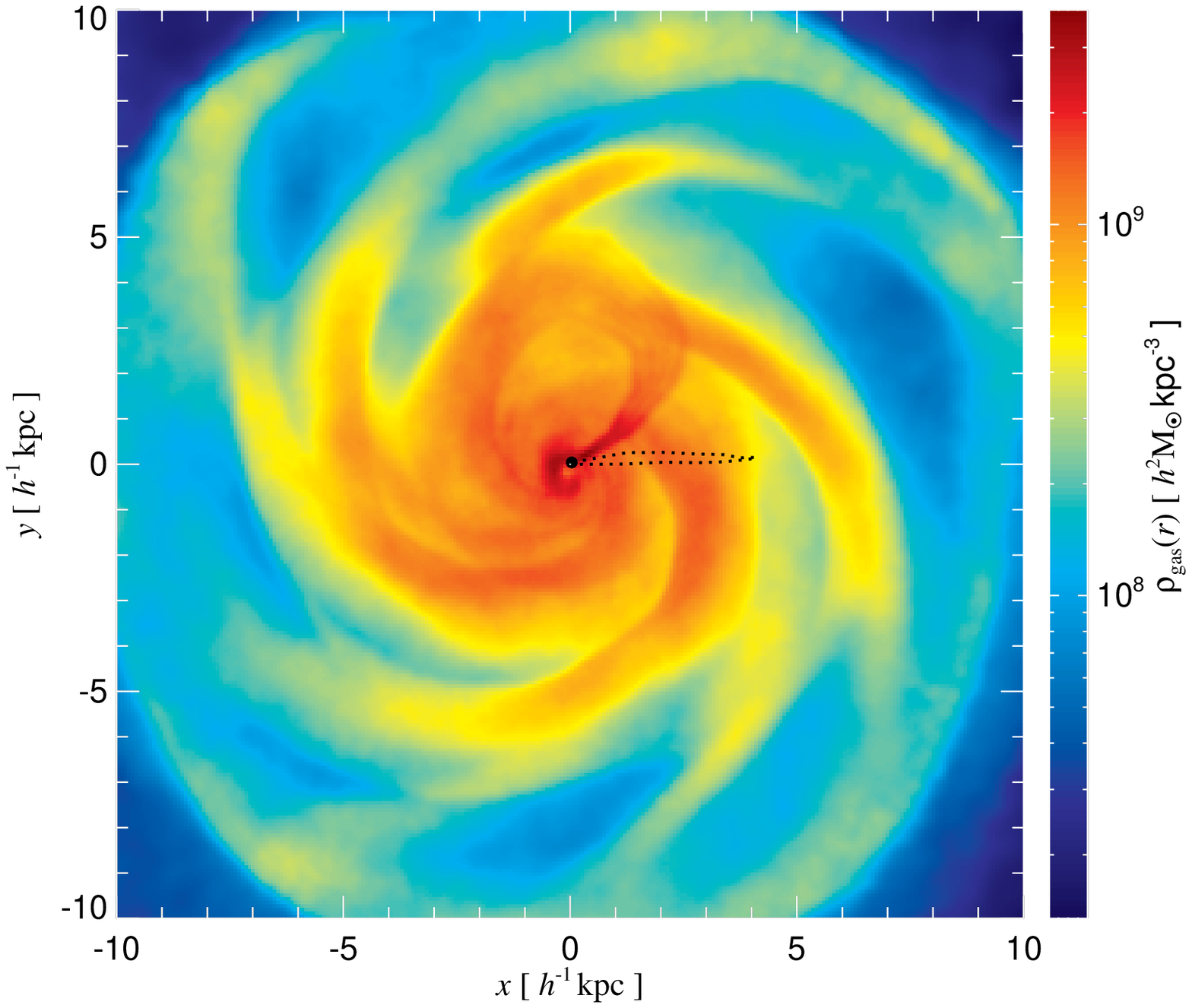}
\hspace{-0.4cm}
\includegraphics[width=6.5truecm,height=6.truecm]{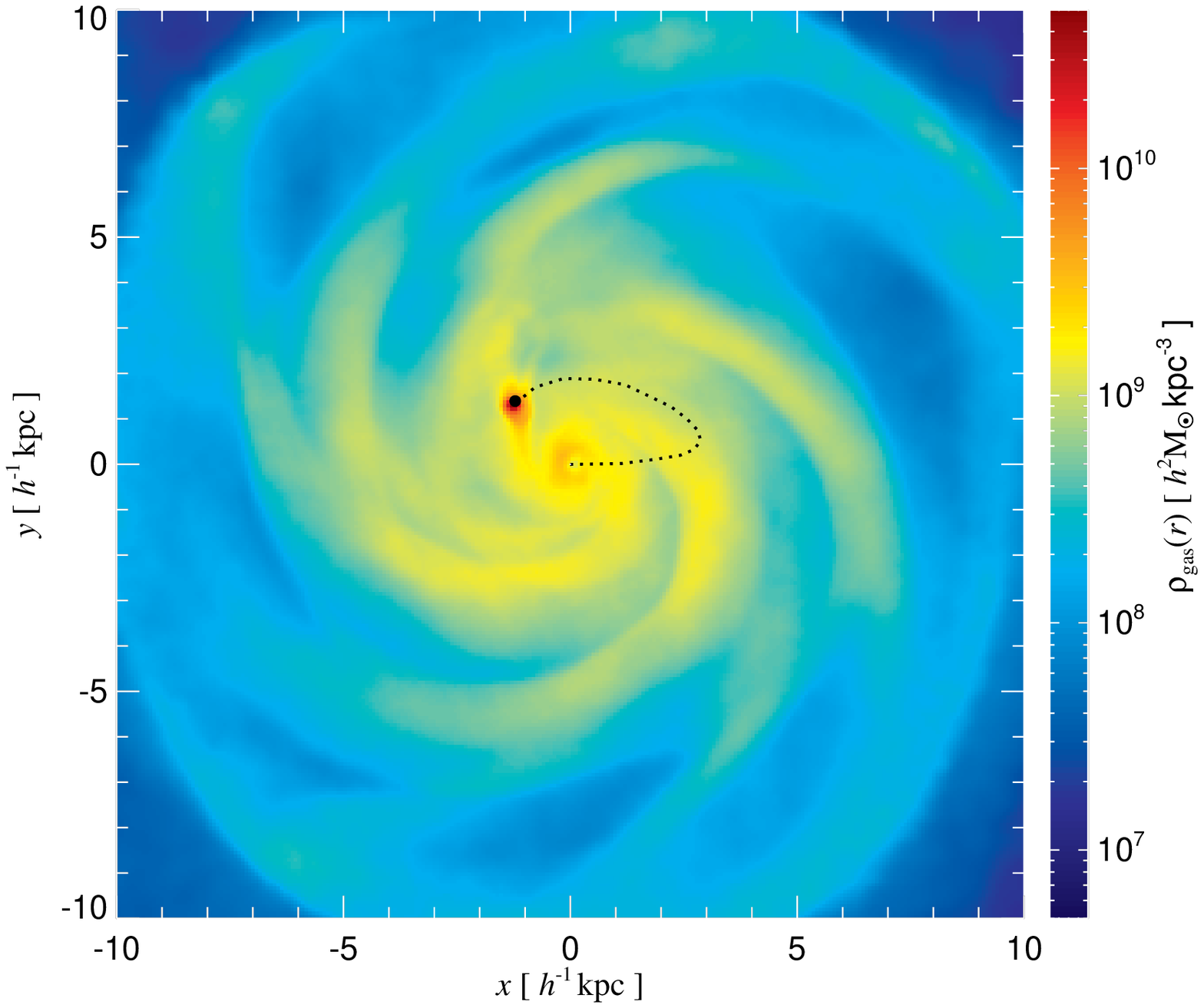}
\hspace{-0.4cm}
\includegraphics[width=6.5truecm,height=6.truecm]{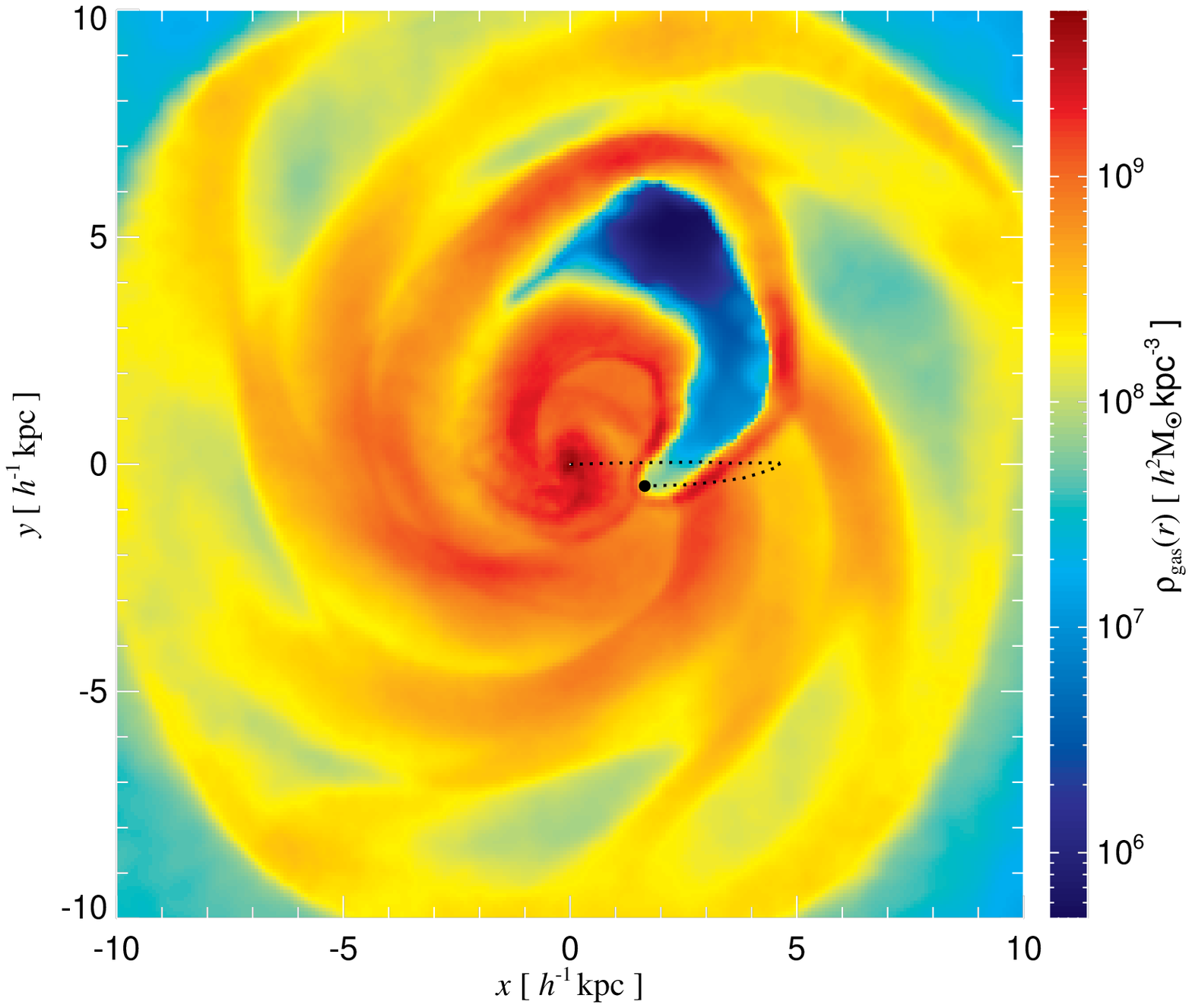}
  }}
\caption{Projected mass-weighted gas density maps of the isolated
  galaxy at time $t=3.92 \times 10^7\,$yrs after BH recoil. The three
  panels show simulation results where the BH does not accrete nor
  exerts feedback (left-hand panel), where the BH accretes but has
  zero feedback efficiency (central panel), and where the BH is
  allowed both to accrete and inject feedback energy (right-hand
  panel). The dotted lines show the BH orbits from the start of the
  simulation. Note that the colour maps display different density
  ranges, as given by the individual colour bars.}
\label{map_iso}
\end{figure*}

\subsection{Galaxy models} 

We first consider an isolated, massive, gas-rich galaxy, that we simulate at
high numerical resolution. The initial conditions of this galaxy are set-up as
described in detail in \citet{Springel2005b}. Specifically, the total mass of
the galaxy is $M_{\rm 200} = 6.28 \times 10^{12}\,h^{-1}{\rm M}_\odot\,$ 
  (with $h = 0.7$), its virial velocity is $v_{\rm 200}= 300\, {\rm km\,
  s^{-1}}$, the dark matter halo concentration is $c\,=\,9$, and the spin
parameter is $\lambda \,=\, 0.033$. The fraction of the total mass in the disc
is $m_{\rm d} \,=\, 0.041$, while the bulge mass fraction amounts to $m_{\rm
  b}\,=\, 0.008$. The galactic disc scale length is $4.8 \,h^{-1} \, {\rm
  kpc}$. The disc is gas-rich, with a gas fraction of $f_{\rm gas}\,=\,0.6$,
while the remaining $40\%$ are in  stars. Circular velocity curves for the
different galactic components are shown in Figure~\ref{vcirc}.

We select the numerical parameters of the multiphase model for star formation
such that the gas consumption timescale is long, specifically we adopt
$t_{*}^{\rm 0} = 12.6\, {\rm Gyrs}$, $A_{\rm 0} = 6000$, $T_{\rm SN}= 6 \times
10^8\, {\rm K}$ \citep[in the notation of][]{SpringelH2003}, and we assume a
softer equation of state with $q_{\rm EOS} \,=\, 0.5$. This choice of
parameters assures that during the simulated time-span (which can be a
considerable fraction of the Hubble time) the galaxy stays comparatively
gas-rich and the gaseous disc is stable. We additionally perform runs where we
keep all numerical and physical parameters the same, except for the equation
of state parameter which we set to 0.05, corresponding to a nearly isothermal
equation of state and resulting in a very clumpy disc. The galaxy is simulated
with $300000$ dark  matter particles in the halo, $200000$ star particles in
the disc, $200000$ gas particles in the disc and $50000$ star particles
belonging to the bulge. The gravitational softening of the disc and bulge is
set to $60\,h^{-1} \, {\rm pc}$, and of the dark matter halo to $2\,h^{-1} \,
{\rm kpc}$.

\begin{figure*}\centerline{\hbox{
\includegraphics[width=10.truecm,height=10truecm]{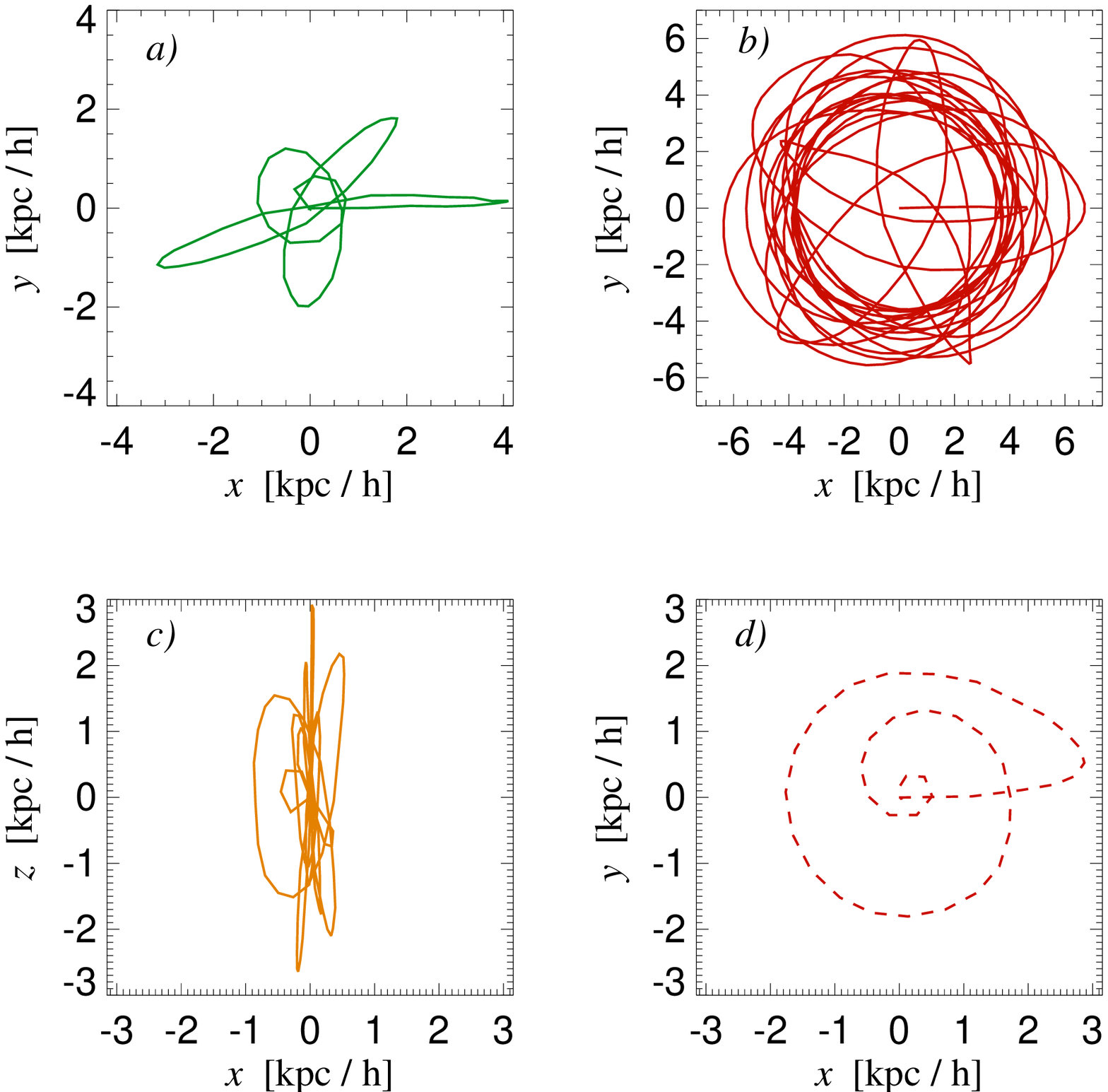}
\hspace{-0.5cm}
\includegraphics[width=8.5truecm,height=8.5truecm]{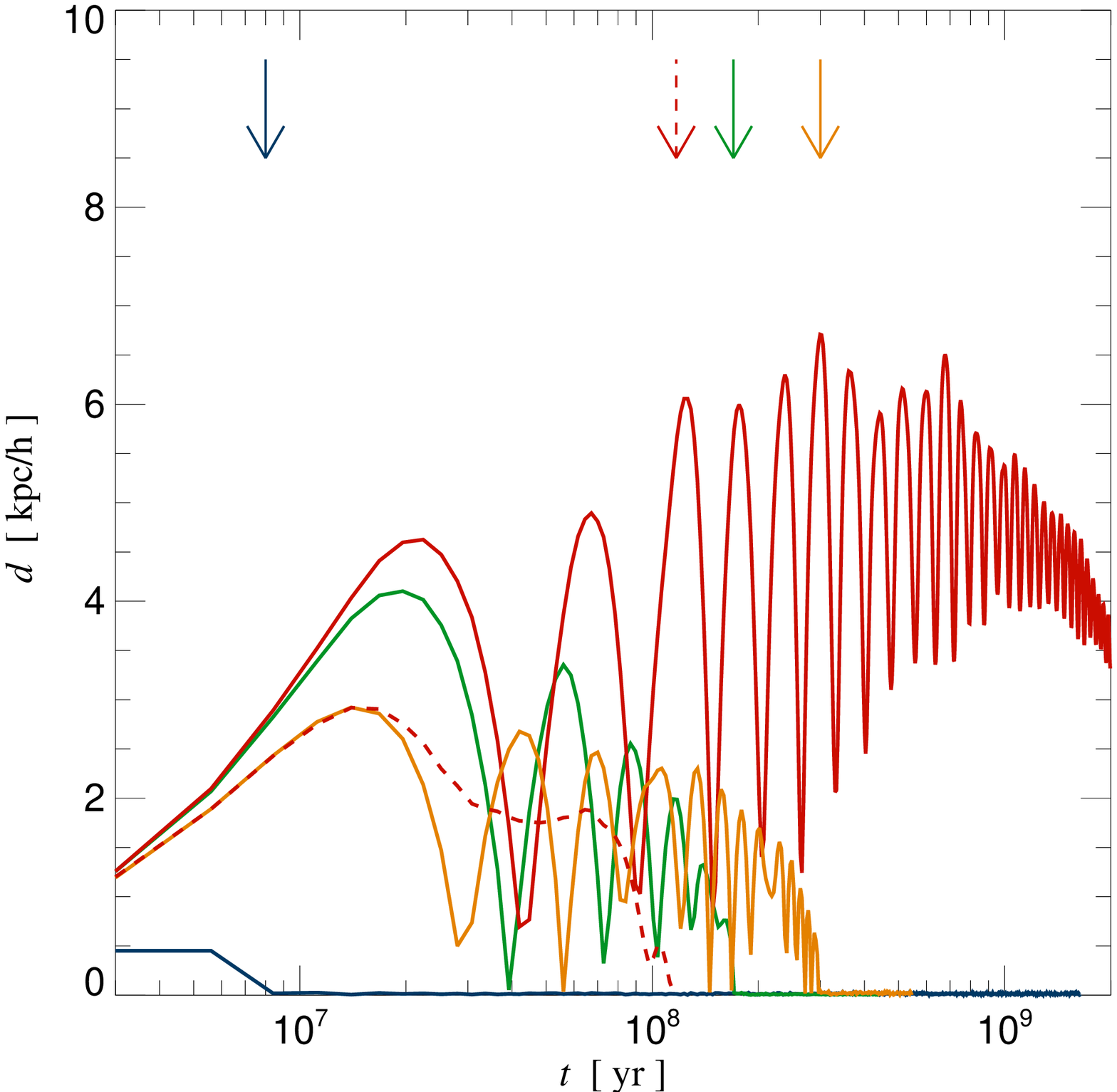}
}}
\caption{{\em Four left-hand panels:} Orbits of BHs over the course of
  the simulated time-span, for different cases. Panel a) $v_{\rm
    kick}= 700 \,{\rm km\, s^{-1}}$ ($\sim 0.5\, v_{\rm esc}$) in the
  galactic plane without BH accretion; Panel b) $v_{\rm kick}= 700 \,
  {\rm km \, s^{-1}}$ in the galactic plane with BH accretion and
  feedback; Panel c) $v_{\rm kick} = 700\, {\rm km\,s^{-1}}$
  perpendicular to the galactic plane with BH accretion and feedback;
  Panel d) $v_{\rm kick} = 700\, {\rm km \, s^{-1}}$ in the galactic
  plane with BH accretion but no feedback. {\em Right-hand panel:} BH
  distance from the minimum of the potential as a function of time for
  the same simulations as in left-hand panels. The additional blue
  line is for an initial recoil of $375 \, {\rm km \, s^{-1}}$ ($\sim
  0.26\, v_{\rm esc}$) in the galactic plane. The arrows indicate when
  the BHs return to the centre of the galaxy.}
\label{BHorbit}
\end{figure*}

\subsection{Numerical setup}

We first evolve the isolated gas-rich galaxy for $2.8 \times 10^8\,$yrs
without a central BH, until the gas disc develops a well defined spiral
structure which is long lived. Thus, during the simulated time of the central
BH the gaseous disc will be in a stable, quasi stationary state. We then
introduce a BH particle in the centre of the galaxy, assuming that such a
gas-rich galaxy should already contain a supermassive BH. To be confident that
the dynamical friction force acting on the BH is numerically well resolved we
set the initial mass of the BH to $5 \times 10^8 \,h^{-1}{\rm M}_\odot\,$ in
most of our calculations\footnote{Note that we have deliberately selected a
  galaxy model with a somewhat smaller bulge mass fraction to simulate the
  long term evolution in a stable state, reducing the occurrence of
  bulge-driven bar instabilities.}. We have also investigated a lower mass BH
of $5 \times 10^7 \,h^{-1}{\rm M}_\odot\,$ (corresponding to $\sim 10^{-3}
M_{\rm bulge}$), and a simulation where the host galaxy is simulated at twice
as high spatial resolution to verify the validity of our findings (see
Appendix~\ref{appen}). Given that we  want to track dynamical friction forces
acting on the BH as accurately as possible, we do not perform simulations with
a lower mass BHs here, because then the mass ratio between the BH and the gas
or star particles would not be sufficiently large. Also, by simulating a more
massive BH we are in a more advantageous regime for spatially resolving
regions close to the Bondi radius.

Once the BH particle is introduced in the centre we let it evolve for an
additional $1.4 \times 10^7\,$yrs\footnote{This delay time is introduced such
  that the BH orbit is not artificially altered by initial accretion caused by
  the introduction of the BH into very gas-rich and dense material.} after
which the BH is imparted a certain recoil velocity, with the direction either
parallel to the galactic disc or perpendicular to it.  Table~\ref{tab_kicks}
summarises the main parameters of our simulated galaxies, as well as initial
BH masses and recoil velocities. For comparison, we also carried out runs
where the BH stays in the centre of the galaxy, and is not subject to any
gravitational recoil. The simulations with the stationary BH permit us to
verify that the BH position always coincides with the minimum of the
gravitational potential within the spatial resolution of the simulation, hence
the resolution is sufficiently high to reduce two-body noise acting on the BH
to negligible levels.

Most of our simulations were performed with BH accretion and feedback
included, but we also considered cases where BH accretion, or selectively only
BH feedback, was switched-off. This allows us to disentangle purely dynamical
processes acting on a recoiled BH from the additional effects due to BH
accretion and feedback, which can affect the properties of the surrounding gas
and thus possibly alter the BH dynamics.

\section{Results}\label{Results}

\subsection{Isolated galaxies}\label{isolated}

\subsubsection{Importance of the stellar and gas dynamical friction}

We first want to assess the importance of dynamical friction forces
acting on the BH as it moves trough the simulated galaxy. For this
purpose, we consider the BH to be a collisionless particle which does
not accrete, and has a constant mass of $5 \times 10^8 \,h^{-1}{\rm
  M}_\odot\,$. Just before the BH is imparted the kick, we estimate
the central escape velocity from the halo based on the minimum of the
gravitational potential, which yields $v_{\rm esc}(0) = \sqrt{2
  \,\Phi(r=0)} \sim 1450\,{\rm km\, s^{-1}}$. Note that the central
escape velocity is rather high, $v_{\rm esc}(0) \sim 4.8\, v_{\rm 200}$,
due to the presence of a central stellar bulge. Using instead the
often adopted equation $v_{\rm esc} = \sqrt{2\, f(c)}\, v_{\rm 200}$,
where $f(c) = c \,(\ln(1+c) - c/(1+c))^{-1}$ is only a function of the
concentration parameter, one would obtain a significantly lower estimate of
$v_{\rm esc}(0) \sim 3.6 \, v_{\rm 200}$, as this considers only
the distribution of the dark matter.

We then impart a gravitational recoil velocity to the BH of $700\,{\rm km\,
  s^{-1}}$ ($\sim 0.5\, v_{\rm esc}$) along the $x$-axis in the plane of the
disc. Figure~\ref{BHdist_drag} shows the resulting distance of the BH from the
minimum of the potential as a function of time for three different runs. The
blue line is for a simulation where the BH particle is massless, and thus is
not subject to dynamical friction forces. The red line is for our default case
where the galactic disc is gas-rich, while the green line is for a simulation
where we have set-up a galaxy with exactly the same structure, only that the
galactic disc is made up entirely of stars. In the case where dynamical
friction is not acting on the BH particle, the apocentric distances reached
during the simulated time-span stay roughly constant. The BH  simple simply
oscillates through the nearly static potential of the galaxy. However, as
expected, dynamical friction from stars, and even more so from the gas, leads
to the systematic reduction of the apocentric distance with time. Dynamical
friction already reduces the first apocentric distance reached, and the effect
becomes more pronounced with further passages of the BH through the innermost
regions. From Figure~\ref{BHdist_drag} we infer that the BH returns to the
centre\footnote{Throughout the paper we define that the BH has returned to the
  centre if its position coincides with the minimum of the potential within
  the gravitational softening length of the simulation. Note that due to the
  limited spatial resolution achieved we cannot track BH-galaxy core
  oscillations \citep{Gualandris2008} which may further prolong BH settling to
  the centre, at least in the case with only stars and no gas at the centre.}
after $\sim 3.3 \times 10^8\,$yrs due to the dynamical friction from a fully
stellar disc. When the host galaxy has instead a substantial fraction of the
disc mass in gas, but otherwise the same structure and the same initial
potential shape, the BH return time is almost halved. The BH trajectory is
sensitive to the composition of the galactic disc in our simulations for three
reasons: i) given that the BH is moving supersonically, dynamical friction is
more efficient if the BH is embedded in a gaseous rather than a stellar
background \citep{Ostriker1999}; ii) unlike stars, gas is collisional and
through radiative cooling it can radiate away energy transferred by the moving
BH and thus form a more concentrated wake behind it, exerting a larger
dynamical friction; iii) radiatively cooling gas is deepening the
gravitational potential with time which also results in shorter return times
to the centre.

\subsubsection{Properties of recoiled BHs}

In Figure~\ref{map_iso}, we show three illustrative examples of recoiled BH
orbiting in the gas-rich galactic disc after $t=3.92 \times 10^7\,$yrs. In all
three simulations, the initial BH mass is $5 \times 10^8 \,h^{-1}{\rm
  M}_\odot\,$ and the recoil velocity is $700\, {\rm km \, s^{-1}} \sim 0.5\,
v_{\rm esc}$ along the $x$-axis. In the left-hand panel we show the case where
the BH is not accreting, in the central panel we present the case where BH
accretion is switched-on but there is no feedback, while in the right-hand
panel we illustrate the case where both BH accretion and feedback are
followed. As can been seen from the left-hand panel, the motion of the BH
through the high density gas creates a density enhancement in its wake 
  \citep{Ostriker1999} that propagates outwards, as has been previously
reported in numerical simulations of a single BH moving through a uniform
  gas cloud \citep{Escala2004}, as well as in simulations of BH binaries in
  gaseous discs during the initial evolutionary phase \citep{Escala2005}. We
further found that the local potential of the stellar disc at the position of
the BH becomes notably deeper, over a typical size of order $\le 500\,h^{-1}
\, {\rm pc}$. The shape of the stellar potential deformation is roughly
spherical, but shows a tendency to be more elongated behind the BH for a
fraction of the time. The reason for a more prominent gaseous wake in
  respect to the stellar wake behind the BH lies in the collisional nature of
  the gas which is allowed to radiatively cool.

When we allow the BH to accrete, the BH orbit is considerably different
(central panel). Even though the BH is moving with a high relative velocity
with respect to the surrounding gas, it does manage to accrete a substantial
amount of gas, allowing it to almost double its mass to $M_{\rm BH}\,=\, 9.1
\times 10^8 \,h^{-1}{\rm M}_\odot\,$ at $t \,=\,3.92 \times 10^7\,$yrs after
the kick. The larger dynamical mass and the considerable drag forces from gas
accretion act in the same direction and prevent the BH from reaching similar
apocentric distances as in the case without accretion. The orbit also
circularises more efficiently in the direction of galaxy rotation when BH
accretion occurs\footnote{While dynamical friction tends to circularise a BH
  orbiting within a galactic disc  -- as it happens in the case where the BH
  is not accreting, see panel (a) of Figure~\ref{BHorbit} \citep[see also][for
    BHs on initially eccentric orbits]{Dotti2007} -- BH accretion can cause
  much more efficient circularisation.}. While initially the accreting BH also
generates a high density wake, after it has gained enough mass some of the gas
particles actually become bound to it and create a high density blob around
the BH.

If in addition to BH accretion we include BH feedback, the BH orbit is again
different (right-hand panel). Instead of a high density wake, we now observe a
low density wake, which similarly propagates outwards from the BH orbit. This
low density wake is due to the BH feedback, which creates a hot expanding gas
plume. Whereas the gas drag from accretion tends to bend the BH orbit in the
prograde direction from the initially radial orbit, the BH orbit is here
deflected in the retrograde direction after reaching the apocentre (opposite
to the rotation of the galaxy). Note also that due to the BH feedback the BH
does not grow much in mass, reaching only $M_{\rm BH}\,=\, 5.1 \times 10^8
\,h^{-1}{\rm M}_\odot\,$ at $t \,=\,3.92 \times 10^7\,$yrs. As a result, the
gas drag from accretion is much smaller than in the previous case. The effect
of the hot, expanding density plume on the evolution of the BH orbit turns out
to be very important, as we discuss in detail below.

In Figure~\ref{BHorbit}, we show the BH orbits and distances from the centre
of the host galaxy during the simulated time-span for four different runs. In
panel (a) the case with $v_{\rm kick}= 700\, {\rm km\, s^{-1}}$ in the
galactic plane without BH accretion is shown. The BH orbit is initially
radial, and after the BH reaches the apocentre, its trajectory is bent in the
prograde direction due to gravitational drag. The BH describes four precessing
loops with decreasing apocentric distances and then quickly spirals inwards
reaching the minimum of the gravitational potential at $\sim 1.7 \times
10^8\,$yrs after the kick. The orbit is essentially completely contained in
the galactic disc, with excursions in the $z$-direction of less than
$100\,h^{-1} \, {\rm pc}$.

When BH accretion but no feedback is included [panel (d)], the BH returns to
the galactic centre on an even shorter timescale, in only $\sim 1.2
\times 10^8\,$yrs. Without self-regulating feedback the BH mass grows rapidly,
making dynamical friction more effective. Moreover, the drag force the BH is
experiencing from accreted gas particles is efficiently circularising the BH
orbit and causing it to co-rotate with the galactic disc. These two processes
acting together lead to rapid spiralling in.

\begin{figure}\centerline{\vbox{
\includegraphics[width=8.5truecm,height=8truecm]{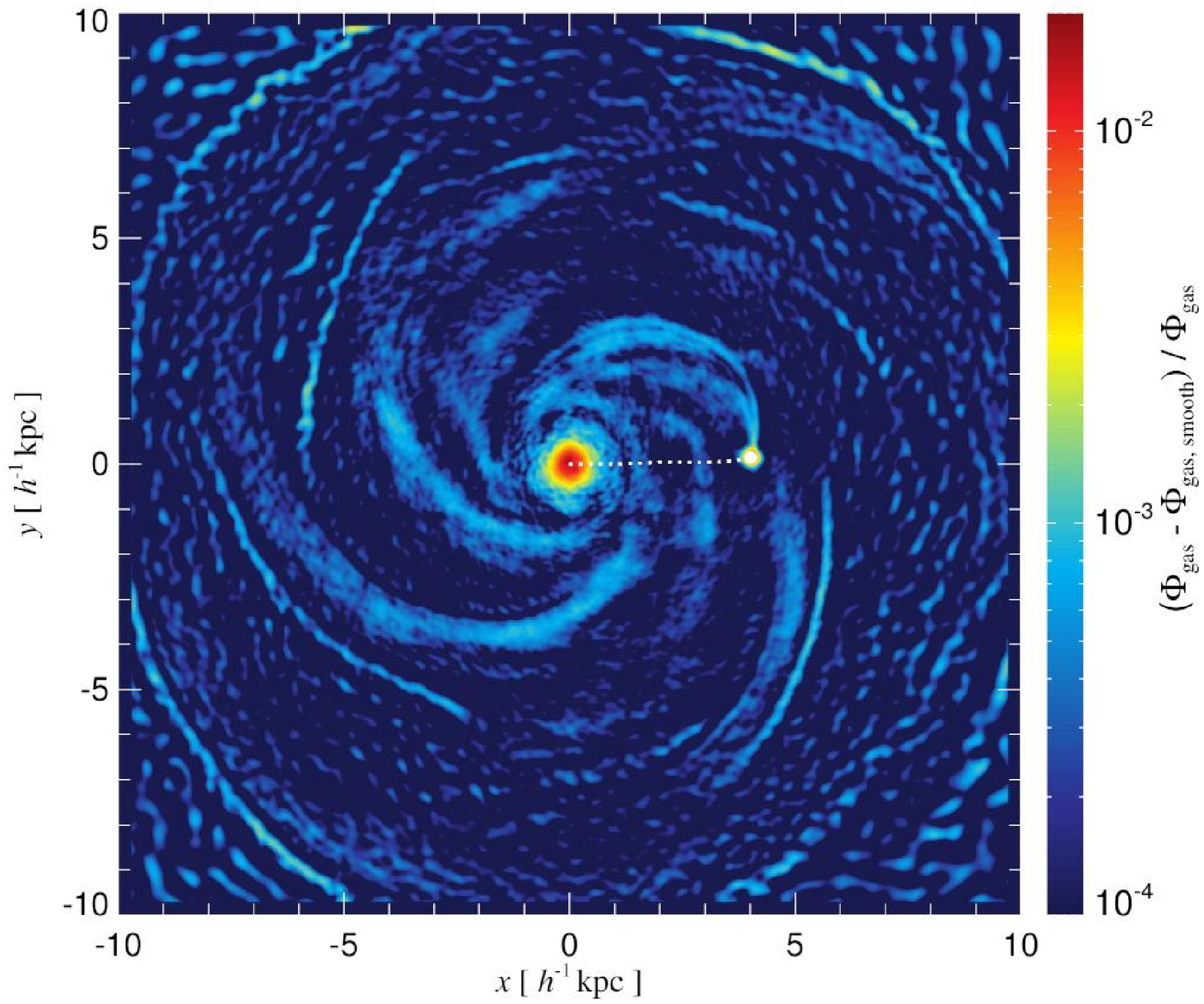}
\vspace{-0.2cm}
\includegraphics[width=8.5truecm,height=8truecm]{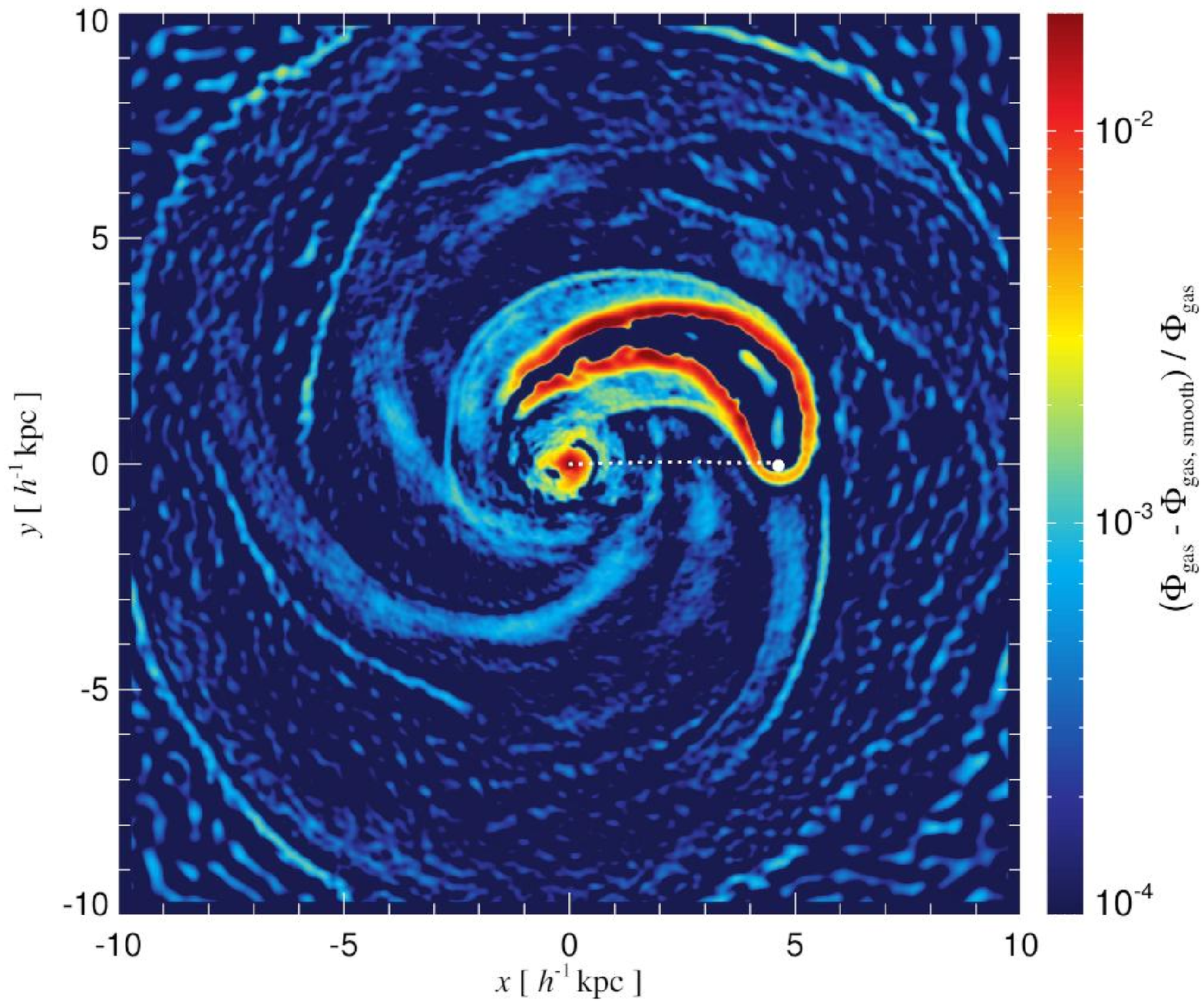}
}}
\caption{Projected unsharp masked map of the gravitational potential
  of the gas (absolute value), for a simulation without BH accretion
  (top panel), and with BH accretion and feedback (bottom panel). The
  displayed snapshots correspond to the moment when the BHs reach the
  apocentres for the first time.}
\label{map_potential}
\end{figure}

BH feedback can nonetheless have an impeding impact on the return timescale of
gravitationally recoiled BHs [panel (b)]. As anticipated above, the BH heats
the surrounding gas, which then forms a hot plume that propagates outwards and
expands, compressing the gas in a thin rim and generating a low density wake
within. To understand in more detail the nature of the resulting BH orbit,
Figure~\ref{map_potential} shows unsharp masked maps of the gravitational
potential of the gas at the moment when the BH reaches the apocentre for the
first time. The unsharp masking has been performed by subtracting from the
initial potential a version smoothed on a scale of $\sim 600\,h^{-1} \, {\rm
  pc}$.  As can be seen in the top panel, for the case without BH accretion,
the high density wake exerts gravitational drag on the BH and the net
gravitational force from the surrounding gas acting on the BH is directed
prograde along the $y$-axis. In the case with BH accretion and feedback, the
situation is more complex, as shown in the bottom panel. Here the BH is
located within the dip associated with high density compressed gas, while
above the BH a hot low density plume is expanding away from the BH. The net
gravitational force acting on the BH from this gas distribution is pointing
retrograde along the $y$-axis, and thus the BH experiences a gravitational
drag in the opposite direction than in the top panel. Moreover, cold gas in
the galactic disc, which approaches the hot gas plume due to the rotation of
the galaxy, is compressed and transfers its angular momentum to the gas in the
plume. This effect further reduces the effective drag force onto the BH from
gas accretion.

The BH describes a precessing elliptical trajectory, and with each passage the
ellipses widen such that after five relatively close passages to the galactic
centre (after about $3 \times 10^8\,$yrs) the orbit starts to circularise, but
with a BH that is counter-rotating, even though it started out with the same
radial motion.  Due to the feedback a ring of hot, low density gas forms,
within which the BH orbits, partially decoupled from the rest of the
galaxy. The BH trajectory is loop-like, with $\pm 500 \,h^{-1} \, {\rm pc}$
excursions from the plane. For the rest of the simulated time of $\sim 1.7
\times 10^9\,$yrs, the BH remains at a considerable distance from the galactic
centre, in the range of $3.5 \,-\, 6.5\,h^{-1} \, {\rm kpc}$, even though
$\sim 7 \times 10^8\,$yrs after the kick, the apocentric distance starts to
decay, indicating that the BH will eventually return to the centre (see also
Appendix~\ref{appen}).

The case where the BH is kicked perpendicular to the plane of the galaxy is
shown in panel (c) of Figure~\ref{BHorbit}. During the first $\sim 3 \times
10^8\,$yrs, the BH describes about a dozen orbits before it returns to the
galactic centre. In this case, the BH feedback does not significantly delay
the return timescale because the BH accretes only during short passages
through the galactic plane, due to the mostly radial orbit along the
$z$-axis. Note also that the maximum distance reached by the BH at the first
apocentre is lower than for the BH kicked in the plane of the galaxy, as a
result of the larger gravitational force exerted from the disc of the galaxy
in this direction (see panel on the right).

Finally, in the right-hand panel of Figure~\ref{BHorbit}, we also show a case
where the kick velocity is much lower, equal to $375\, {\rm km\,s^{-1}}$ (blue
line). Here the BH briefly leaves the minimum of the potential and reaches a
radial distance of $500 \,h^{-1} \, {\rm pc}$, but then returns to the centre
within $7 \times 10^6\,$yrs.

\begin{figure}
\includegraphics[width=7.5truecm,height=20truecm]{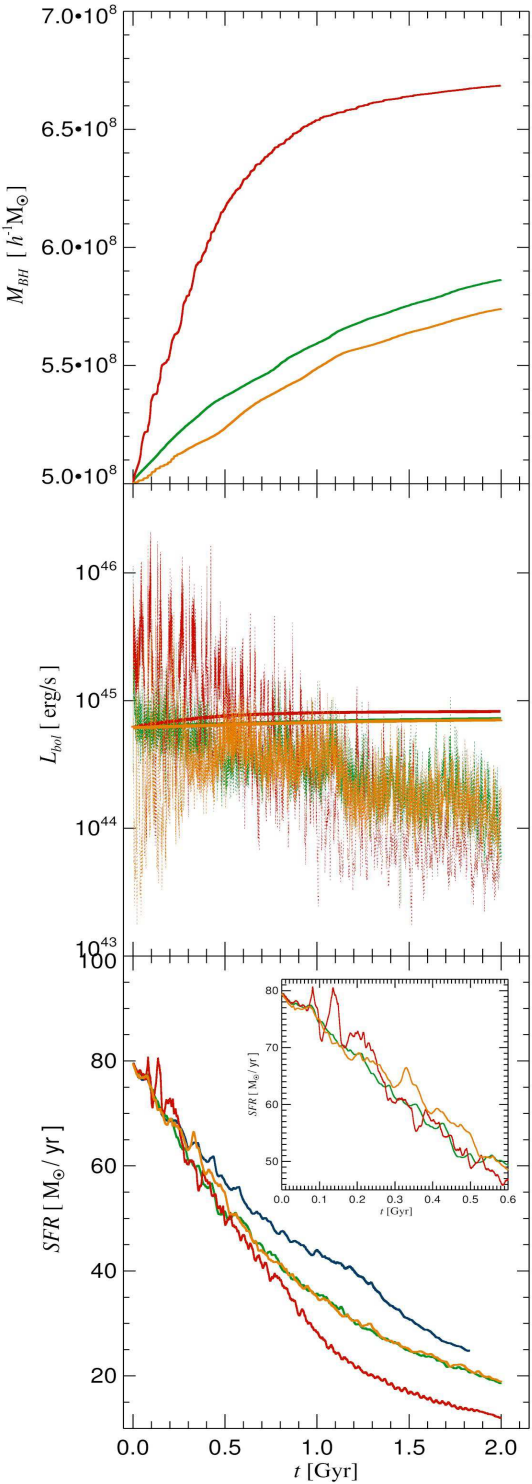}
\caption{BH mass (top panel), bolometric luminosity (central panel) and total
  SFR (bottom panel) as a function of time. The green lines are for the run where
the BH is always at the centre of the galaxy. Cases where the BH experiences
a gravitational recoil of $700\, {\rm km\,s^{-1}}$ in the plane of the galaxy or
perpendicular to the galactic plane are denoted with red and orange lines,
respectively. The blue line (only in the bottom panel) is for the simulation
without a BH. Continuous lines in the middle panel denote an accretion rate
equal to $0.01$ of the Eddington rate.}
\label{mbh_iso}
\end{figure}

In Figure~\ref{mbh_iso} we show the BH mass (top panel) and bolometric
luminosity (middle panel) as a function of time, as well as the time evolution
of the total star formation rate (SFR) of the host galaxy (bottom panel) for
different cases. The green lines are for the case where the BH does not
experience a gravitational recoil, while the red and orange lines are for
$v_{\rm kick}= 700\, {\rm km \, s^{-1}}$ parallel and perpendicular to the
plane of the galaxy, respectively. From the top panel it can be seen that the
BH which does not experience any recoil increases its mass moderately, by
$\sim 8.5 \times 10^7 \,h^{-1}{\rm M}_\odot\,$ over $2\,$Gyrs. This is due to
secular processes which gradually transport some gas towards the inner
regions, fuelling BH accretion. The bolometric luminosity\footnote{For the
  computation of the bolometric luminosity a radiative efficiency of $0.1$ is
  assumed throughout the paper.}  is initially around $10^{45}\,{\rm
  erg\,s^{-1}}$ and slowly decreases with time, but we note that after $\sim
0.5\,$Gyrs the BH accretion falls below $0.01$ of the Eddington rate, as
indicated by the continuous lines (becoming probably radiatively
inefficient). Thus, for most of the time this AGN would be optically dim,
exhibiting only very few, brief luminous episodes.

In contrast, the BH kicked in the plane of the galaxy initially grows more as
it encounters a larger supply of material on its orbit through the
gas-rich galactic disc. Consequently, its bolometric luminosity is up to an
order of magnitude higher than that of the BH which never leaves the
centre. Once the BH orbit circularises in a ring of low density material
formed by feedback, its accretion is even more sub-Eddington than that of the
BH which stays in the galactic centre. Note however that the AGN bolometric
luminosity obtained in this case should be an upper limit for
several reasons. First, it is probably not very common that the BH is
gravitationally recoiled exactly within the disc, as assumed here, given that
at present there is no evidence for a correlation between the spin
orientations of the BH and of the host galaxy. Second, during a galaxy merging
event, which is a much more realistic setting for the occurrence of a
gravitationally recoiled BH, the largest amount of gas available for accretion
will be in central regions, meaning that a kicked BH will be biased towards
accreting less gas (see Section~\ref{merging}). Finally, the BH accretion rate
estimated from equation~(\ref{Bondi_eq}) should be considered as an upper
limit if the gas surrounding the BH is not multiphase, and the
BH feedback is not strong enough to self-regulate the BH growth. While for a
stationary BH in the centre of the host galaxy this is unlikely to occur, for
a recoiled BH the actual accretion rate may well be lower if it 
leaves the dense multiphase medium. We explore this issue in detail in
Appendix~\ref{appen}. Nonetheless, our findings suggest that the recoiled AGN
could have accretion rates up to a few percent of the Eddington rate on
timescales of a few $10^7$ yrs, if their orbits are approximately contained within
the gas-rich galactic disc.

Gravitational recoil of the BH perpendicular to the galactic disc
significantly suppresses BH accretion, but it does not truncate it all
together. As the BH orbit decays towards the centre, the accretion rate
increases as the BH experiences more and more passages through the
disc. Eventually, once back in the centre the accretion rate is very similar
to the case of the stationary BH, and the difference between the final masses
is not very large, i.e. $\sim 10^7 \,h^{-1}{\rm M}_\odot\,$. This is, however,
very likely a lower limit to the mass difference between a recoiled and a
stationary BH, given the quiescent nature of the host galaxy. In a more
realistic scenario, where the progenitor BHs merge during a merger of two
galaxies, a large amount of gas will be funnelled towards the central
regions. This gas will form a copious reservoir for BH accretion and it will
thus make a much bigger difference for BH growth whether the remnant BH stays
in the centre or is gravitationally recoiled, as we discuss in
Section~\ref{merging}.

\begin{figure*}\centerline{
\hbox{
\includegraphics[width=8.truecm,height=14truecm]{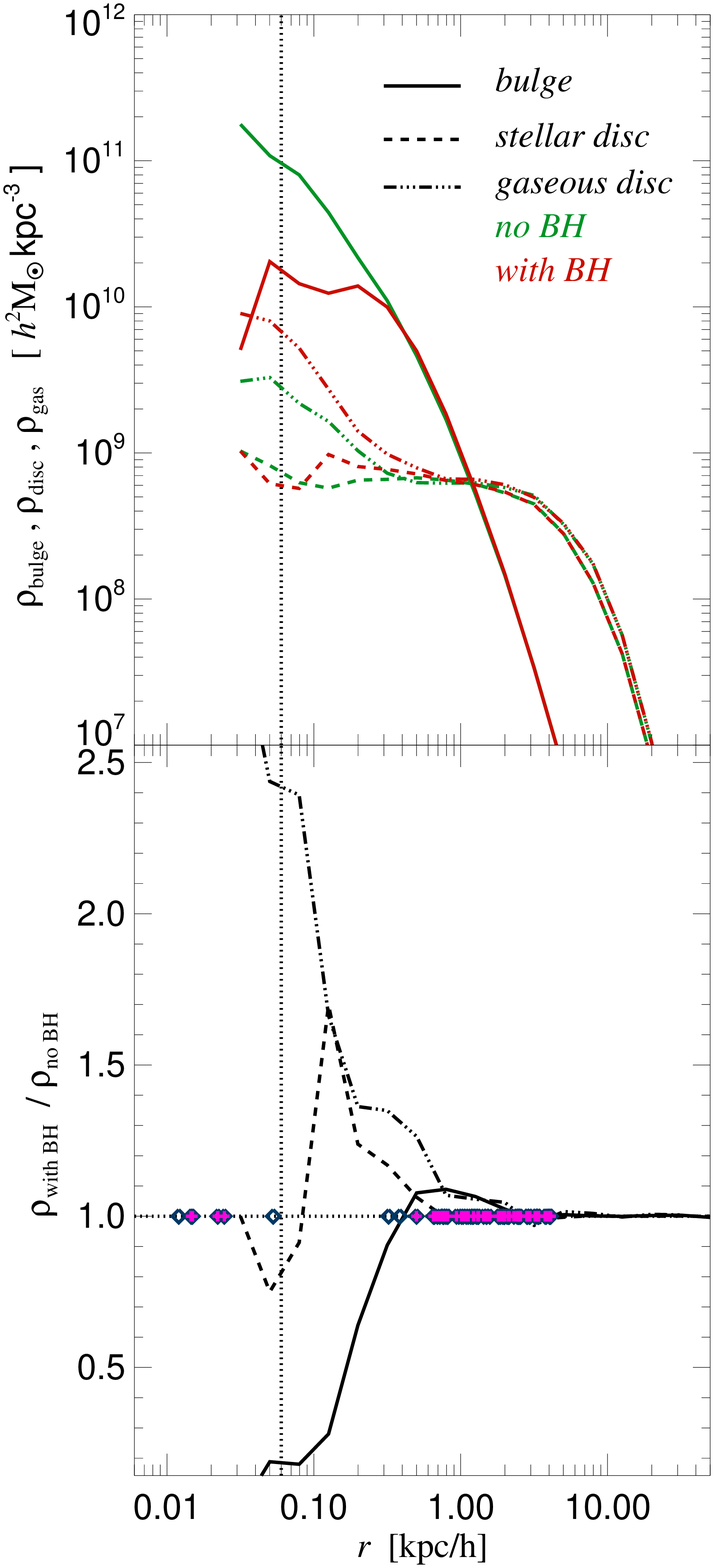}
\includegraphics[width=8.truecm,height=14truecm]{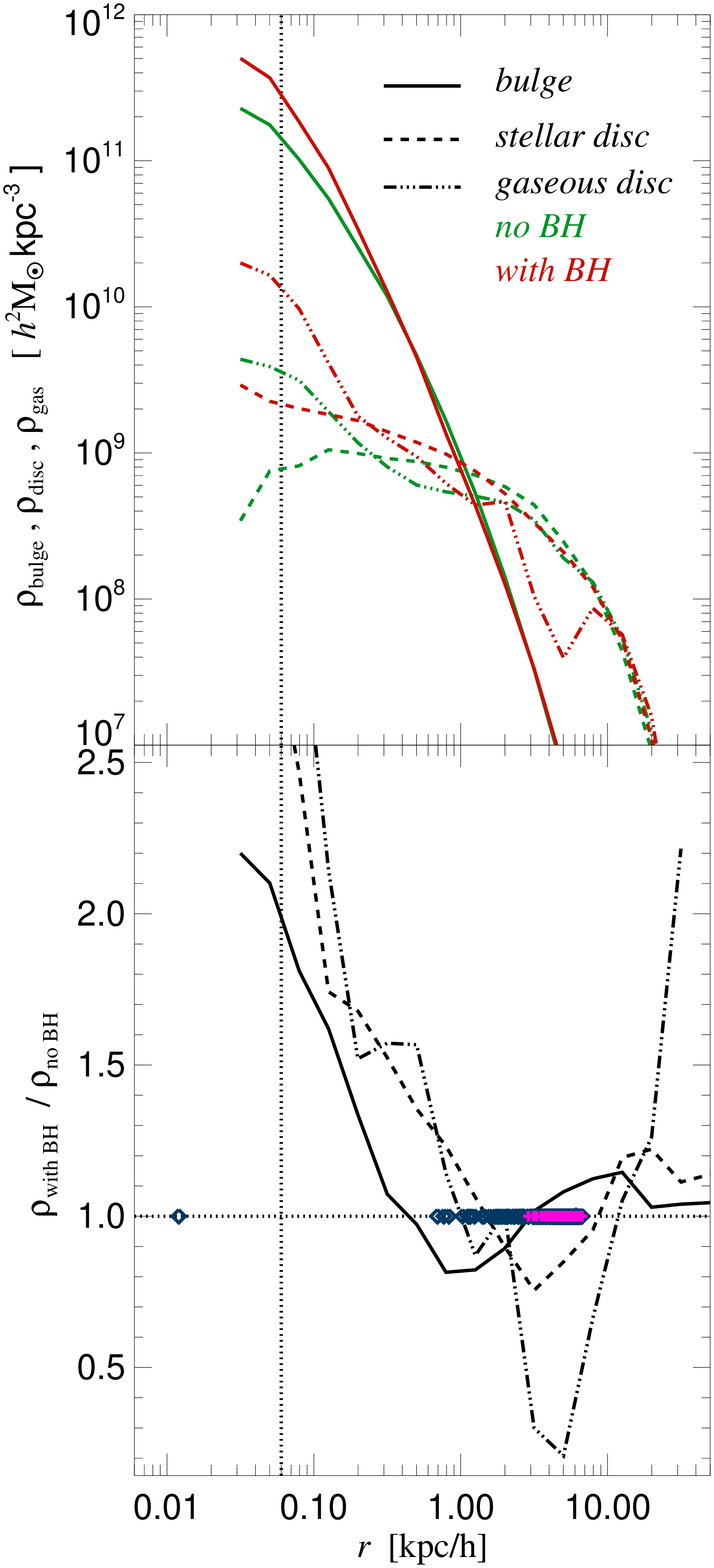}
}}
\caption{Mass density profiles of the stellar bulge (continuous
  lines), stellar disc (dashed lines) and gaseous disc (triple
  dot-dashed lines) are shown in the upper panels for simulations with
  recoiled BHs (red colour) and without BHs (green colour). The ratio
  of these density profiles for each galactic component (same line
  styles) is plotted in the bottom panels. The BH is subject to an
  initial recoil of $700\, {\rm km \, s^{-1}}$ in the galactic
  plane. Results for a BH without (with) gas accretion are shown in
  the left-hand panel (right-hand panel). Blue diamond symbols denote
  the BH distance from the minimum of the potential, while the pink
  crosses indicate those distances where the BH velocity is less than
  $400\, {\rm km \, s^{-1}}$.}
\label{bulgeprop}
\end{figure*}

\subsubsection{Impact of recoiled BH on the host galaxy}

In the bottom panel of Figure~\ref{mbh_iso}, we show the total SFR of the
simulated galaxy, where the blue line denotes the simulation result without a
BH, for comparison. The feedback from the stationary BH reduces the SFR of the
host galaxy in the central regions during the simulated time span. In the case
where the BH undergoes a gravitational recoil perpendicular to the galactic
disc, the SFR of the host galaxy is similarly diminished as in the case where
the BH is stationary, because the recoiled BH returns relatively quickly back
to the centre. On the other hand, a BH which is kicked in the plane of the
galaxy is found to have a sizable impact on the amount of stars produced, both
because the BH then grows more in mass (feedback effects are hence stronger) and
also because the BH can affect larger portions of the star-forming disc on its
orbit.

We do however find that the motion of the BH through the galaxy triggers brief
local bursts of star formation. These are highlighted in the inset plot, where
we show the evolution of the SFR during the first $0.6\,$Gyrs after the
kick. Feedback from recoiling BHs generates expanding plumes of hot gas, which
push the surrounding material and induce localised bursts of star formation
with rates of several ${\rm M}_\odot\,{\rm yr}^{-1}$.

In Figure~\ref{bulgeprop}, we show radial density profiles of stars in the
bulge (continuous lines), stars in the disc (dashed lines) and gas in the disc
(triple dot-dashed lines). The results for the simulations without BH and for
the gravitationally recoiled BH are denoted with green and red colour,
respectively. In the left-hand panels we illustrate the case where the
recoiled BH was not allowed to accrete, and the profiles are computed once the
BH has returned to the centre of the galaxy. During the simulated time-span,
the density of stars in the bulge decreases systematically in the case with
recoiled BH, as can be seen more clearly in the lower panel, where the ratio
of the profiles is shown. This result agrees well with those from N-body
simulations of recoiled BHs in stellar cores \citep[see][and references
  therein]{Gualandris2008}, where the BH scatters the surrounding stars
transferring some of its kinetic energy. In the particular case studied here,
the central bulge mass deficit (evaluated within $1\,h^{-1} \, 
{\rm kpc}$) is of the order of the BH mass once the BH returns to the
centre. Moreover, we find that the stars in the galactic disc are also
perturbed by the orbiting BH, but there is no clear systematic trend with the
radial distance from the centre as a function of time. Interestingly, however,
the high density gas in the wake of the BH follows the BH all the way to the
centre, creating a large density enhancement.

In the right-hand panels of Figure~\ref{bulgeprop}, we show density profiles
for the case where BH accretion and feedback have been switched-on. Here, as
discussed above, the BH does not return to the centre, but instead describes a
loop-like orbit, counter-rotating with respect to the disc. The BH spends
considerable amount of time away from the central regions, as indicated by the
coloured symbols, and transfers kinetic energy to the stars which are not in
the centre. This causes a clear dip in the density of stars in the bulge for
$0.5-3\,h^{-1} \, {\rm kpc}$ (see bottom panel). Stars which are
scattered-away from this region create density enhancements on both sides,
towards the centre and away from it. Thus, instead of a mass deficit the bulge
mass within $1\,h^{-1} \, {\rm kpc}$ is increased by $\sim  0.7 \times M_{\rm
  BH}$. In this case there is also a systematic reduction of the density of
stars in the galactic disc in the region of the disc where the BH spends
considerable amount of time (pink symbols).  The bulge density declines more
steeply than the disc density, and thus for radii $\ge 1\,h^{-1} \, {\rm kpc}$
the BH is more likely to transfer its kinetic energy to the stars in the
disc. Additionally, due to the BH feedback in this region a ring of hot, low
density gas creates a local perturbation in the gravitational potential, which
is felt by the stars in the disc. A dip in the density distribution of the
disc stars is formed.

While we have seen that the recoiling BH which returns to the centre
(left-hand panels of Figure~\ref{bulgeprop}) leads to a moderate mass deficit
of the bulge, of the order of the BH mass, in some cases the mass deficit can
be substantially bigger. In the simulation where the BH is recoiled
perpendicular to the galactic disc, and passes many more times close to the
galactic centre, the mass deficit becomes of the order of $\sim 3.5 \times
M_{\rm BH}$ when the BH finally returns to the centre.

\subsubsection{Clumpy discs}\label{clumpy}

\begin{figure}\centerline{
\includegraphics[width=8.5truecm,height=8truecm]{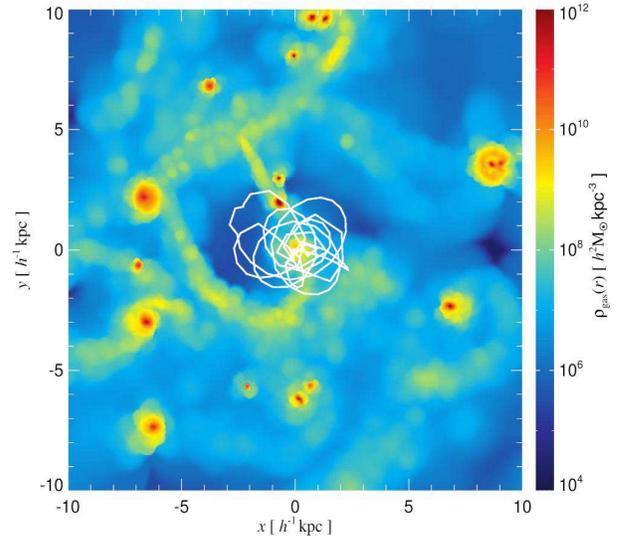}}
\caption{Projected mass-weighted gas density map of the isolated galaxy with
  a very clumpy disc. The over-plotted white line shows the BH orbit from the beginning
  of the simulation to the moment when it returns to the galaxy centre.}
\label{isogal_eos0.05}
\end{figure}

At present the spatial distribution of stars and gas in high redshift galaxies
are still rather uncertain, yet this is the regime where BH mergers and thus
gravitational recoils should be most frequent. Recent observational studies
suggest that there is a significant population of high redshift galaxies that
have gas-rich, thick and clumpy discs \citep[][and references
  therein]{Forster2009}. To numerically explore this scenario, we now discuss
cases where the galactic disc is very clumpy instead of stable and
quasi-stationary.

Even though in the model we consider the total galaxy mass and virial radius
are the same, the clumpy disc has a deeper central potential
with a central escape velocity $v_{\rm esc} \sim 1980\, {\rm km\, s^{-1}}
\sim 6.6\,v_{\rm 200}$, as a result of more efficient radiative cooling in the
innermost regions. While in the case of a smooth disc a BH with kick velocity
$\ge 0.3\, v_{\rm esc}$ can leave the innermost regions, here a larger kick
velocity of $ > 0.5\, v_{\rm esc}$ is needed to displace the BH
from the central region at all. Also, the maximum distance reached at the
first apocentre is affected by the structure of the gaseous disc: if the disc
is relatively smooth, a BH recoiled with $ \sim 0.5\, v_{\rm esc}$ reaches
$\sim 4.6 \,h^{-1} \, {\rm kpc}$, while if the disc is clumpy a BH with
initial kick velocity of $ \sim 0.8\, v_{\rm esc}$ does not reach a distance
farther than $3\,h^{-1} \, {\rm kpc}$ from the centre. To the extent that
clumpy discs are accompanied by elevated deposition rates of baryons in the
centre, as it happens in our simulation, BHs are more likely to stay in the
centre, or leave only for brief periods of time, when they experience merger
kicks.  This may contribute to explain the very scarce observational
evidence for off-centred quasars \citep{Bonning2007}.

An example of a gravitationally recoiled BH with $v_{\rm kick} = 1600 \,{\rm
  km\,s^{-1}}$ ($ \sim 0.8\, v_{\rm esc}$) in the plane of the clumpy disc is
shown in Figure~\ref{isogal_eos0.05}. The BH orbit (denoted with a white line)
is plotted on top of the projected density map, where the clumpy nature of the
disc can be clearly seen. Regardless of gas accretion and feedback processes,
the BH returns to the minimum of the potential in $3 \times 10^8\,$yrs. The BH
trajectory is completely contained in the disc, with $|z| < 100 \,h^{-1} \,
{\rm pc}$, and soon after reaching the first apocentre the BH starts
co-rotating with the galactic disc. During its orbit through the clumpy disc,
the BH accretes a relatively small amount of gas. By the time the BH returns
to the centre, the BH mass has increased by $\sim 3.6 \times 10^7 \,h^{-1}{\rm
  M}_\odot$. Most of the mass gain happens when the BH is on its way back to
the innermost regions, passing through a dense central lump of gas. The
bolometric luminosity of the BH moving through a clumpy disc is on average
lower than that of the BH passing though a more uniform gas distribution: in
fact, the bolometric luminosity is reduced by at least one order
of magnitude with respect to the values shown in Figure~\ref{mbh_iso}. During
most of the $3 \times 10^8\,$yrs while displaced from the centre, the BH
should be in the radiatively inefficient accretion regime with only a few
brief bursts of typical duration $< 10^7\,$yrs when passing through a dense
gas lump.

When the BH finally approaches the centre, its accretion rate increases,
making the AGN optically bright for $\sim 5 \times 10^7\,$yrs, after which the
accretion rate drops again due to ensuing feedback. This suggests an
interesting possibility for spectroscopically detecting a recoiled BH on its
way back to the centre of a galaxy, given that its relative velocity during
the luminosity maximum is still sufficiently large ($\sim 500 {\rm
  km\,s^{-1}}$) to discriminate it from a stationary BH. However, it is
possible that the BH accretes the gas with a significant time delay compared
to our model estimates, in which case it might become optically bright only
when it has already settled into the minimum of the potential.

In the case of a clumpy disc we do not see any clear evidence for the impact
of BH feedback on the surrounding gas, e.g.~in the form of a low density wake
in the gas as we witnessed for the smooth disc. We conclude that while BH
feedback can alter the local gas density  distribution and thus influence the
BH dynamics, these feedback effects are not universal and depend sensitively
on the thermodynamic state and the spatial distribution of the local gas.

\subsection{Merging galaxies}\label{merging}

The simulations presented in Section~\ref{isolated} have been instructive for
understanding the complex interaction of gravitationally recoiled BHs with
host galaxies that have pure stellar disc or gas-rich discs with different
equations of state. However, these models considered isolated, unperturbed
systems, which is not very realistic for the most commonly expected BH
mergers, and this simplification may hide important aspects of the evolution
of the recoiled BHs.

In order to illustrate the changes expected in a more realistic case, we now
consider a simulation of the major merger of two equal mass galaxies. The two
merging galaxies have been set-up initially in exactly the same way as
described for the isolated gas-rich galaxies, with each of them containing a
central BH with a mass of $5 \times 10^7 \,h^{-1}{\rm M}_\odot$. The galaxies
collide on a prograde parabolic orbit, break due to dynamical friction of
their dark matter halos, and eventually coalesce to form a spheroidal remnant
system. When the galaxy cores merge, their central BHs merge as well, leading
to the recoil of the BH merger remnant.

\begin{figure*}\centerline{\vbox{
\hbox{\includegraphics[width=7.5truecm,height=7truecm]{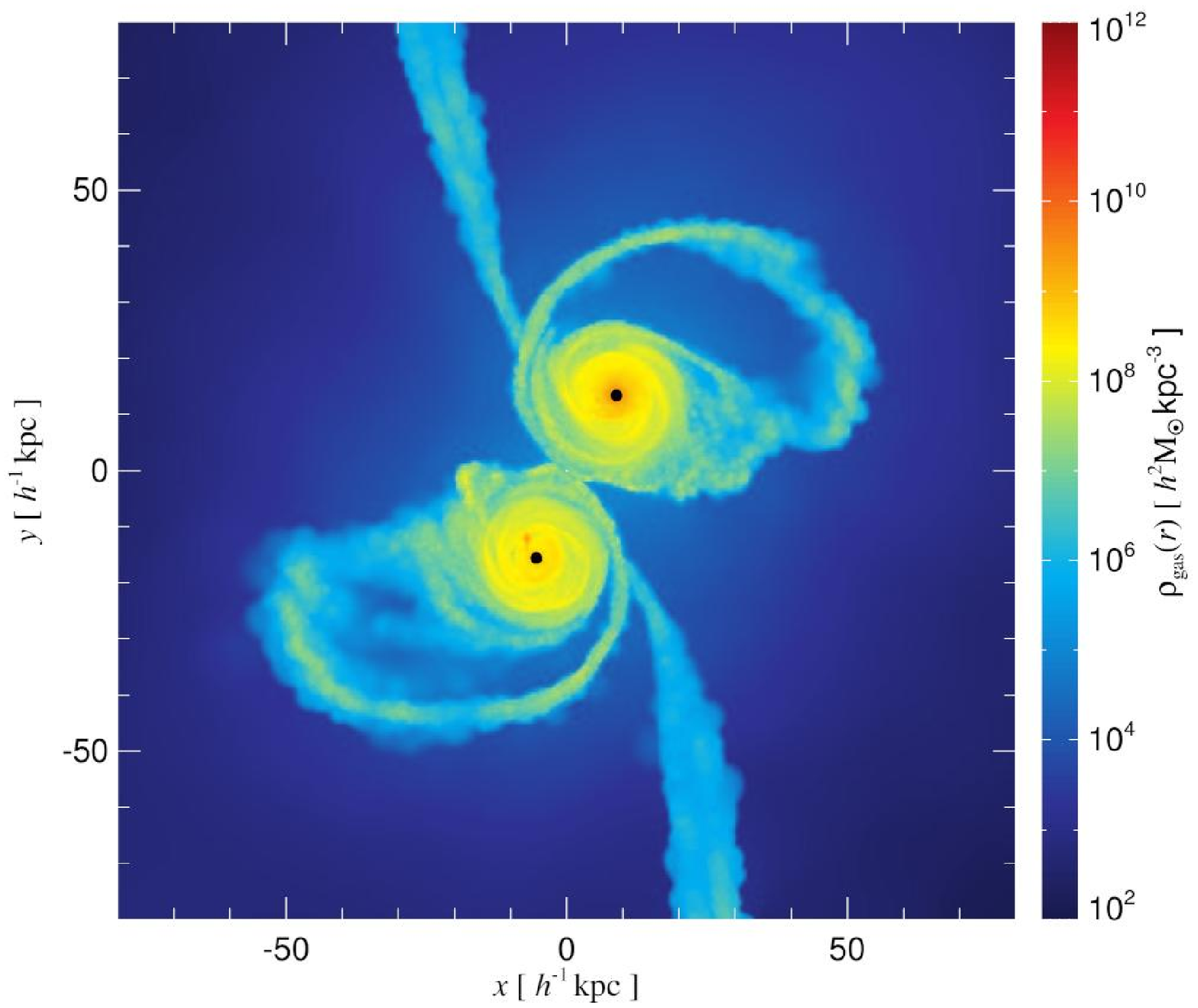}}
\vspace{-0.5cm}
\hbox{\includegraphics[width=7.5truecm,height=7truecm]{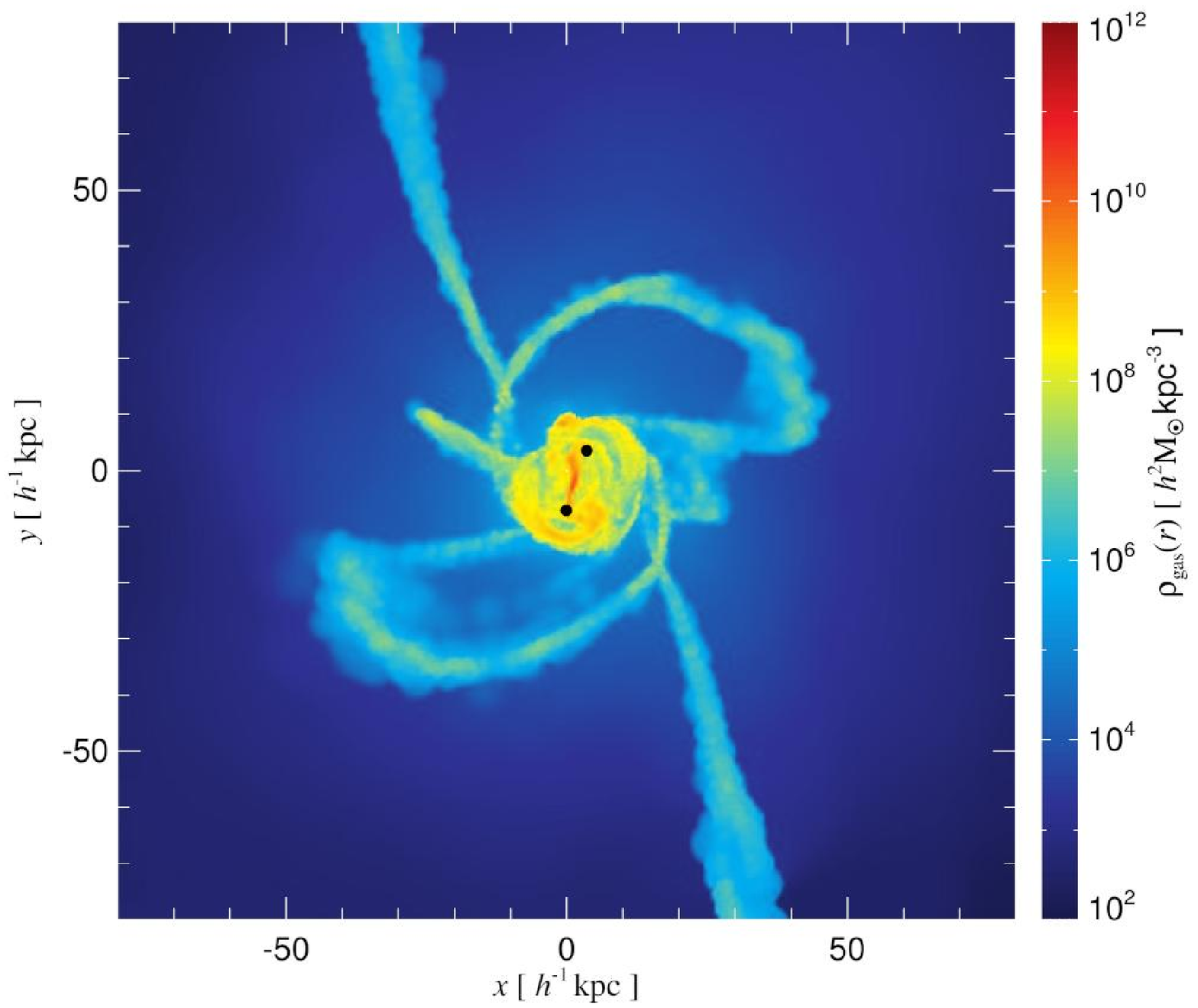}}
\vspace{-0.5cm}
\hbox{\includegraphics[width=7.5truecm,height=7truecm]{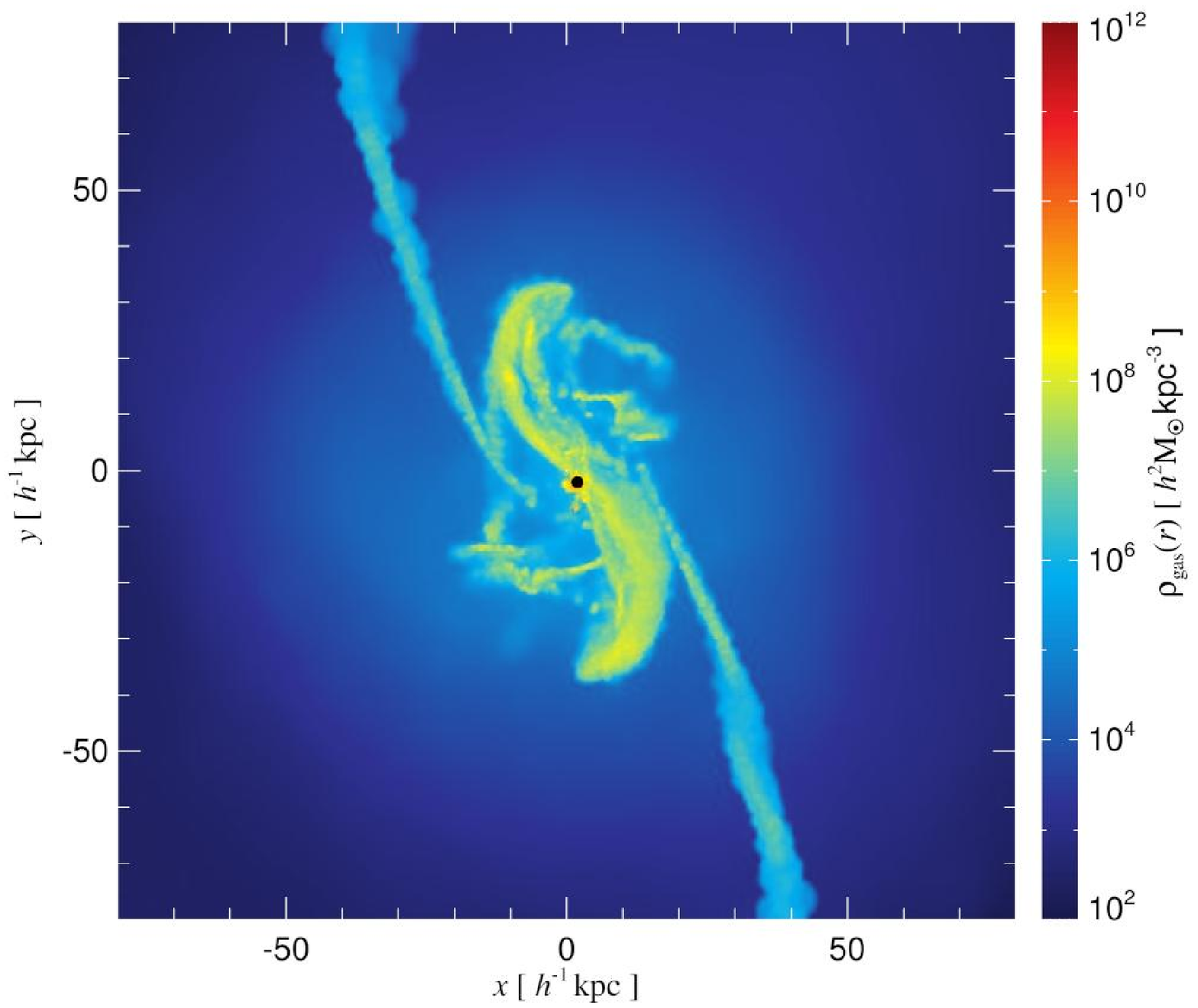}}}
\vbox{
\vspace{-1cm}
\hbox{\includegraphics[width=10.5truecm,height=10truecm]{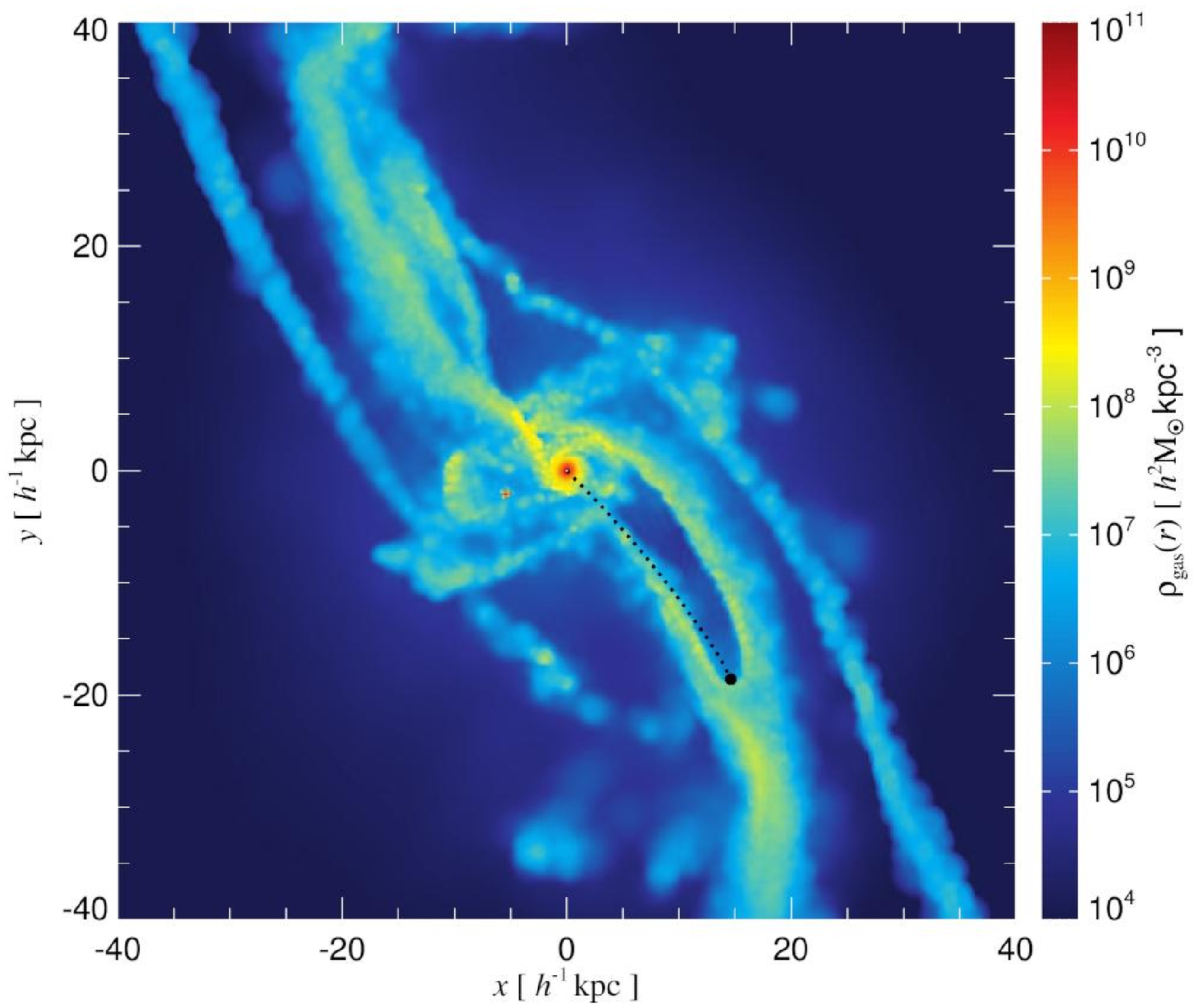}}
\hbox{\includegraphics[width=10.5truecm,height=10truecm]{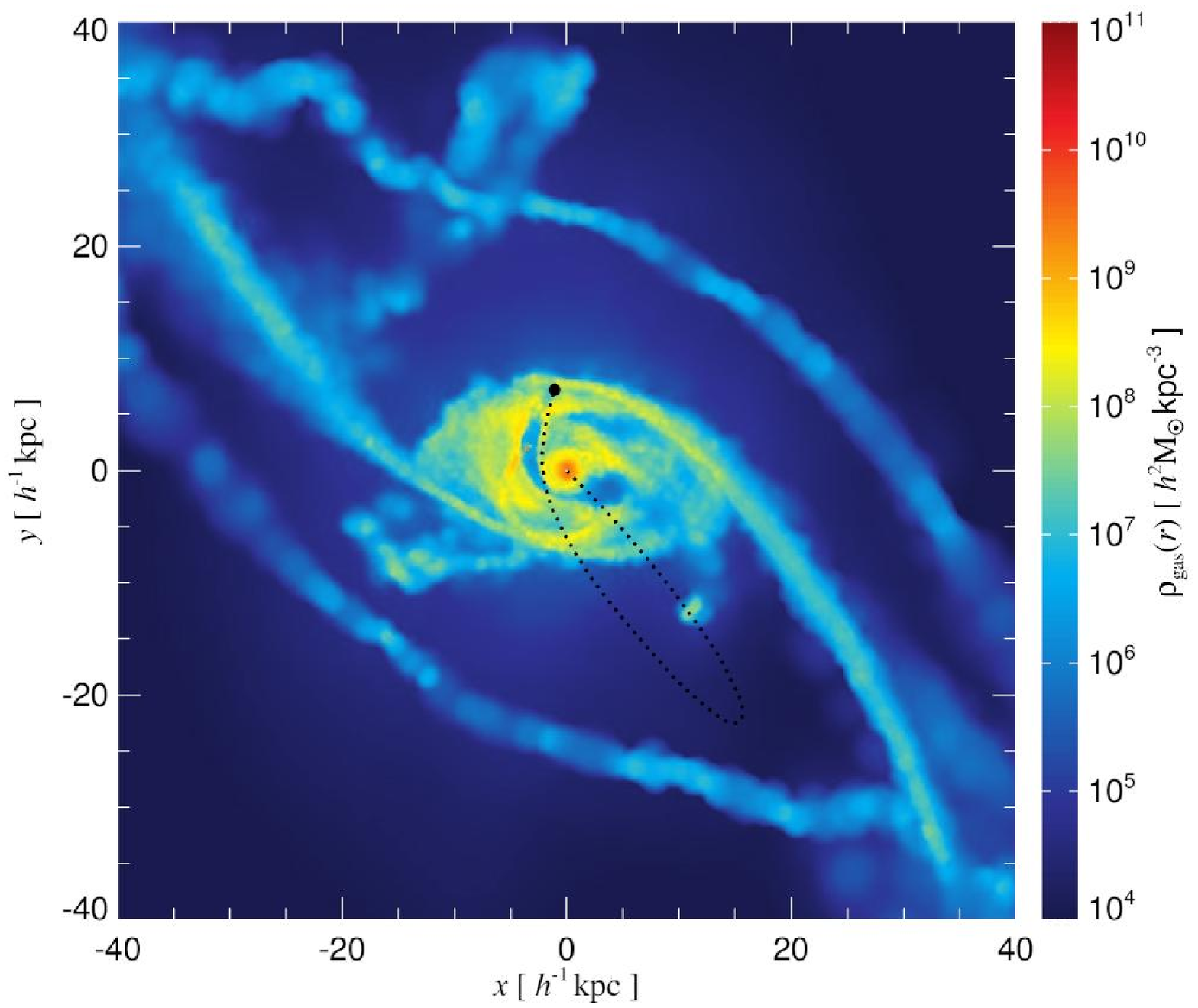}}}}
\caption{Projected mass-weighted gas density maps of two gas-rich
  merging galaxies, each containing a supermassive BH in the
  centre. The left-hand panels show a time sequence at $t=1.55$,
  $1.62$ and $1.69\,$Gyrs from the beginning of the simulation. At
  $1.69\,$Gyrs, the two BHs coalesce. In the right-hand panels we show
  the orbit of the gravitationally recoiled BH at times $t=1.75$ and
  $1.89\,$Gyrs, which at the moment of the merger was kicked with a
  recoil velocity of $2500 \, {\rm km \, s^{-1}}$ along the $x$-axis
  and $-2500\,{\rm km\,s^{-1}}$ along the $y$-axis.}
\label{combgal}
\end{figure*}  

In Figure~\ref{combgal}, we show the time evolution of the gas density during
the major merger of the two gas-rich galaxies. The left-hand panels correspond
to times prior to the BH coalescence, while in the right-hand panels we
illustrate snapshots after the remnant BH has been gravitationally kicked with
$v_{\rm kick}\sim 3535 \, {\rm km\,s^{-1}}$. In this particular case, we
deliberately selected an initial BH trajectory that should maximise the
possible interaction with the dense gaseous arms which are falling towards the
core. The top right-hand panel shows a wake of low density, hot gas behind the
BH, similarly to the case of the isolated galaxy with a smooth disc. However,
the occurrence of feedback-generated hot gas is short lived and is present for
only $\sim 5\times 10^7\,$yrs, as a result of the rapidly changing gas
distribution in the significantly disturbed system. After reaching the
apocentre, the BH passes trough the dense disc-like structure which is forming
around the galactic core (bottom right-hand panel). Here BH feedback leads
again to the formation of a low density wake. Thus, if BH accretion and
feedback processes are not significantly suppressed compared to our numerical
model, and if the gas discs are sufficiently smooth, such wakes of hot, low
density gas may be ubiquitous in post merger systems, albeit short-lived.

\begin{figure}\centerline{
\includegraphics[width=8.5truecm,height=8truecm]{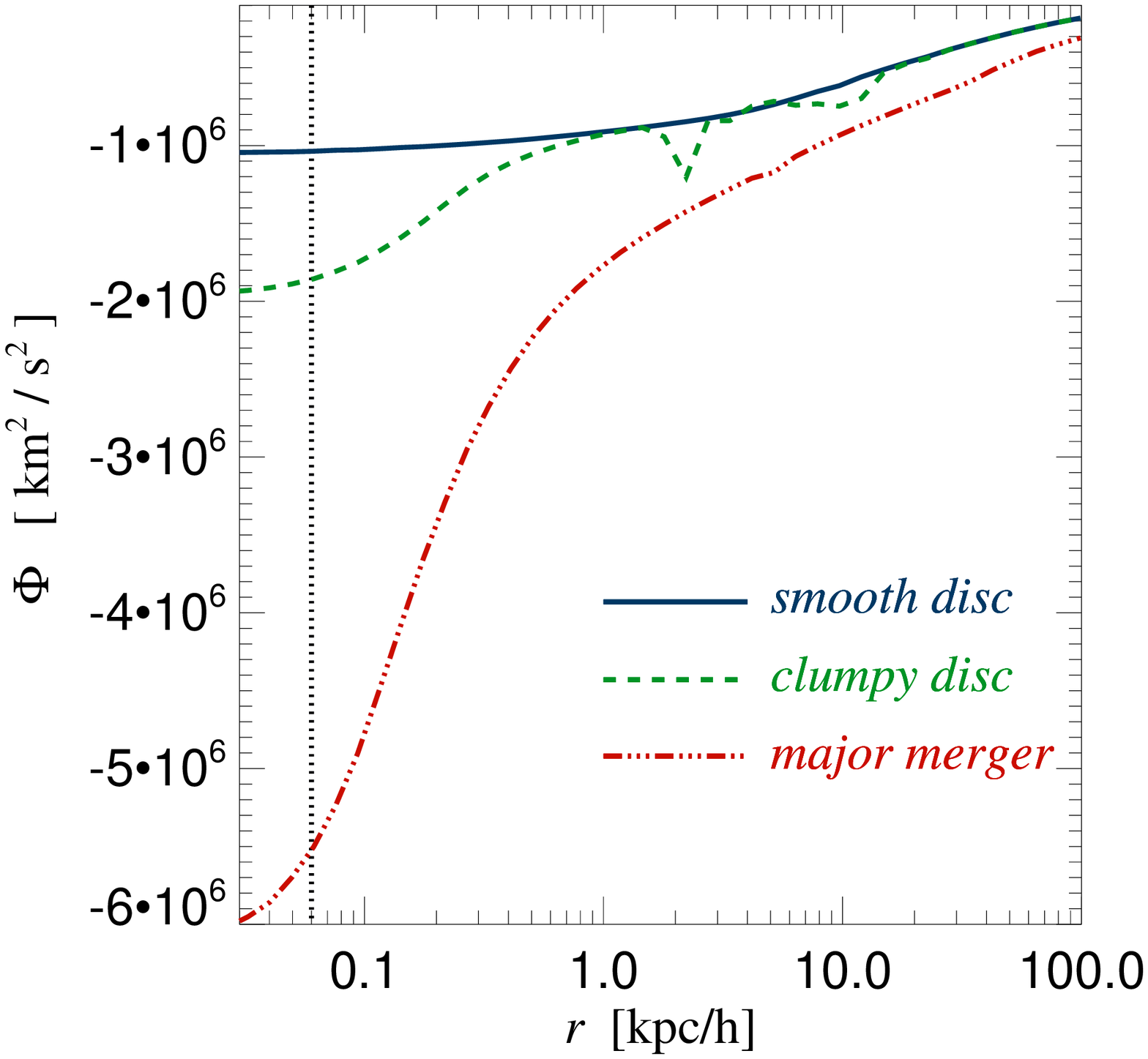}}
\caption{Radial profiles of the gravitational potential for the isolated
  galaxy with a smooth disc (blue continuous line), for the isolated galaxy with a
clumpy disc (green dashed line), and for the system undergoing a major merger
of two gas-rich galaxies (red triple dot-dashed line). The potential in the
central regions is progressively both deeper and steeper in these three cases.}
\label{potential}
\end{figure}

We note that the recoil velocity of $v_{\rm kick}\sim 3535\,{\rm km\,s^{-1}}$
imparted on the remnant BH is extremely high. At present it is unclear whether
such kicks are attainable \citep{Campanelli2007, Baker2008}, but it seems
certain that they should be very rare. The estimated central escape velocity
from our merging system at the moment of BH coalescence is $v_{\rm esc} \sim
3510\,{\rm km \, s^{-1}}$. This is also a strikingly large value, caused here
by the fact that we consider comparably massive galaxies in which substantial
baryonic dissipation has created a massive central stellar bulge.  For recoil
velocities $\le 0.8\,v_{\rm esc}$, the remnant BH does not leave the innermost
regions, making it highly unlikely for the BH to ever get kicked out from the
centre. To illustrate why recoil velocities $\le 0.8\,v_{\rm esc}$ are not
sufficient for the BH to leave the galactic core, in Figure~\ref{potential} we
show radial profiles of the gravitational potential for our merging system as
well as for isolated galaxies with smooth and clumpy discs, for
comparison. Clearly, already in the case of the clumpy disc, the gravitational
potential is not only almost twice as deep as the gravitational potential of
the galaxy with a smooth disc, but it is also much steeper in the inner parts,
which explains why the BH needs to have a larger fraction of the central
escape velocity to leave the innermost regions. For the merging system, the
central potential is almost six times deeper. Much of this depth arises in the
steep part within $r < 1 \,h^{-1} \, {\rm kpc}$, so that the BH indeed has to
have a recoil velocity comparable with the escape speed to leave the
core. Similar results have been found for a variety of gas-rich mergers,
extending to lower mass systems \citet{Blecha2011}.

In order to explore the imprints of strong gravitational recoils in this
system, we hence consider very high kick velocities: $v_{\rm kick} \sim 3253\,
{\rm km \,s^{-1}} \sim 0.9\,v_{\rm esc}$ and $v_{\rm kick} \sim 3535\, {\rm km
  \,s^{-1}} \sim v_{\rm esc}$, which could be achieved if the BHs are close to
maximally spinning, are of comparable mass, and have anti-aligned spins in the
orbital plane, as we assume here. Note that the probability of such a high
recoil velocity is very low for a typical merger, highlighting that even
maximal recoil velocities currently proposed might not be sufficient for
expelling the BHs from massive, gas-rich hosts. Nonetheless, by adopting these
very high kick velocities we can explore the properties of recoiled BHs in a
realistic merging setting and draw firm qualitative conclusions without loss
of generality. The BH orbits for the two kick velocities turn out to be rather
different due to the shape of the gravitational potential. For $v_{\rm kick}
\sim 0.9\,v_{\rm esc}$ the BH reaches only $1.5 \,h^{-1} \, {\rm kpc}$ at the
first apocentre. After experiencing several mostly radial-type passages close
to the centre, it returns to the centre in $10^8\,$yrs. Instead, the BH kicked with
$v_{\rm kick} \sim v_{\rm esc}$ leaves the central region all together. Along
its way towards the outskirts it experiences drag forces from the infalling
dense gaseous arms, such that it only reaches an apocentric distance of $\sim
27 \,h^{-1} \, {\rm kpc}$. Thereafter, the BH describes a precessing
elliptical trajectory, and after reaching each apocentre, the BH moves
retrograde with respect to the forming circum-nuclear disc (similar to the
simulation findings in the case of the uniform isolated disc). This will
likely increase the time needed for the BH to return to the centre. In fact,
at time $\sim 10^9\,$yrs after the recoil event the apocentric distances are
still very large, $\ge 20 \,h^{-1} \, {\rm kpc}$, implying that this BH will
be wandering within the host galaxy for several Gyrs before returning to the
centre.

\begin{figure*}\centerline{\hbox{
\includegraphics[width=6.2truecm,height=5.8truecm]{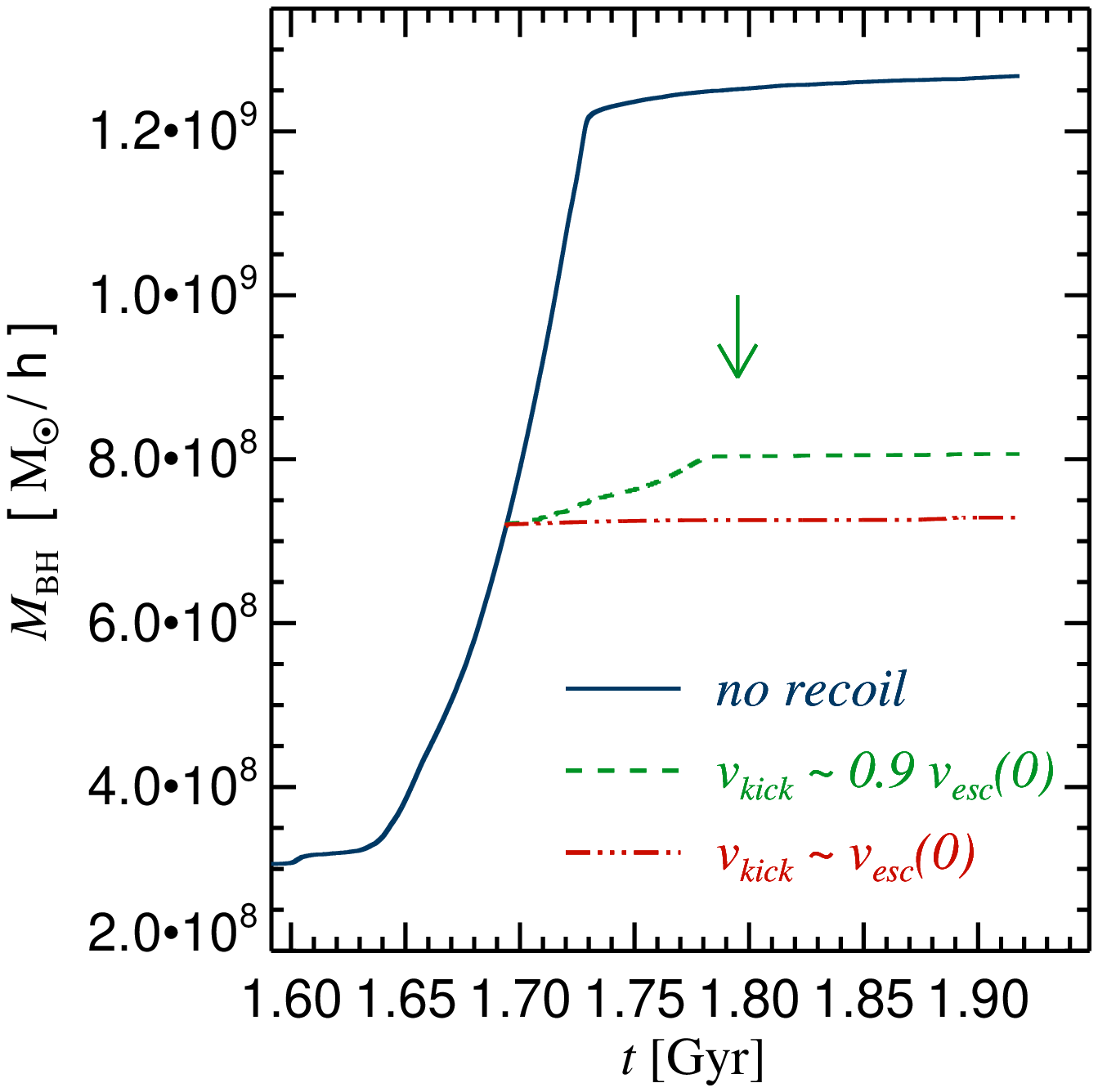}
\hspace{-0.35cm}
\includegraphics[width=6.2truecm,height=5.8truecm]{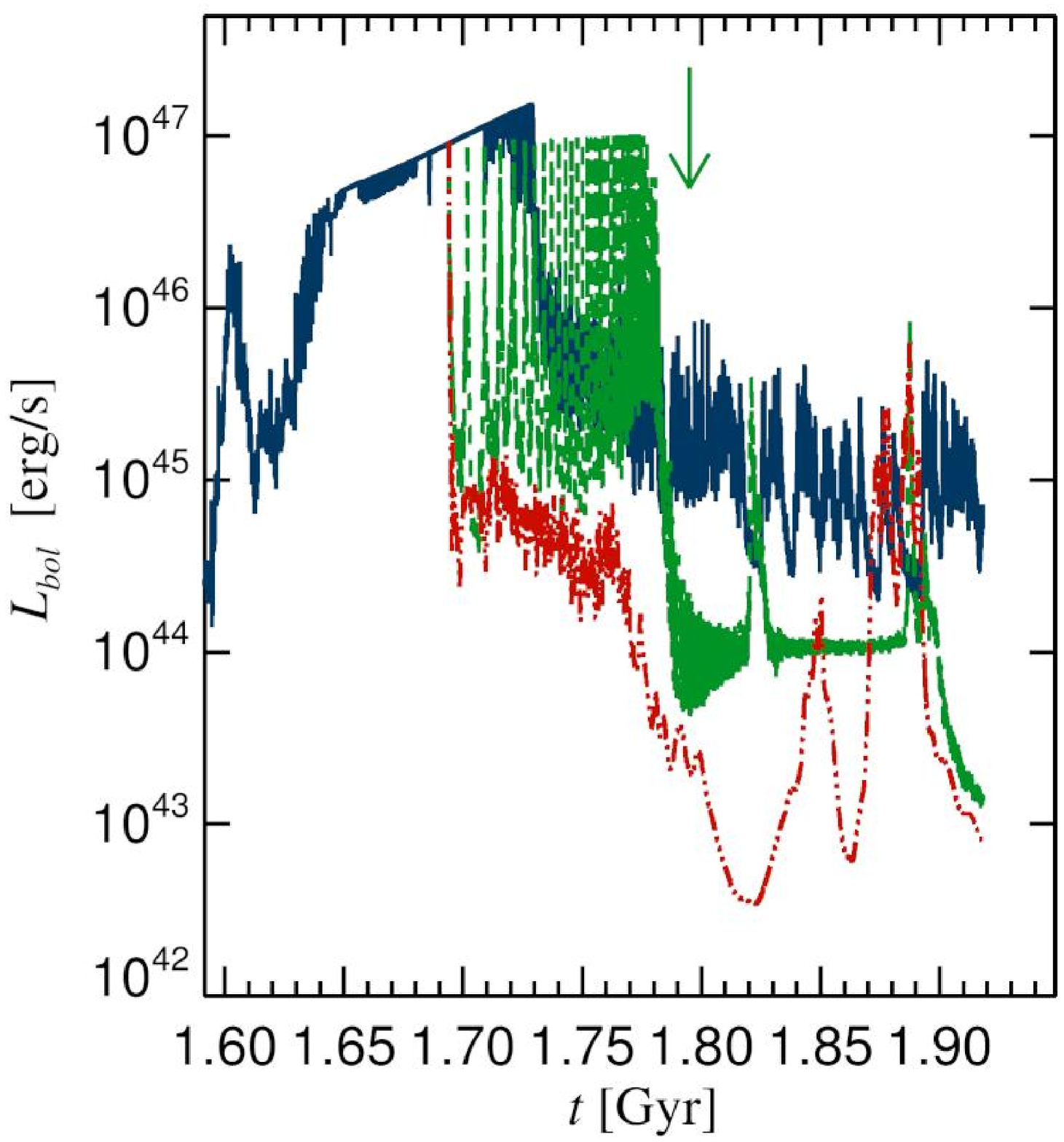}
\hspace{-0.35cm}
\includegraphics[width=6.2truecm,height=5.8truecm]{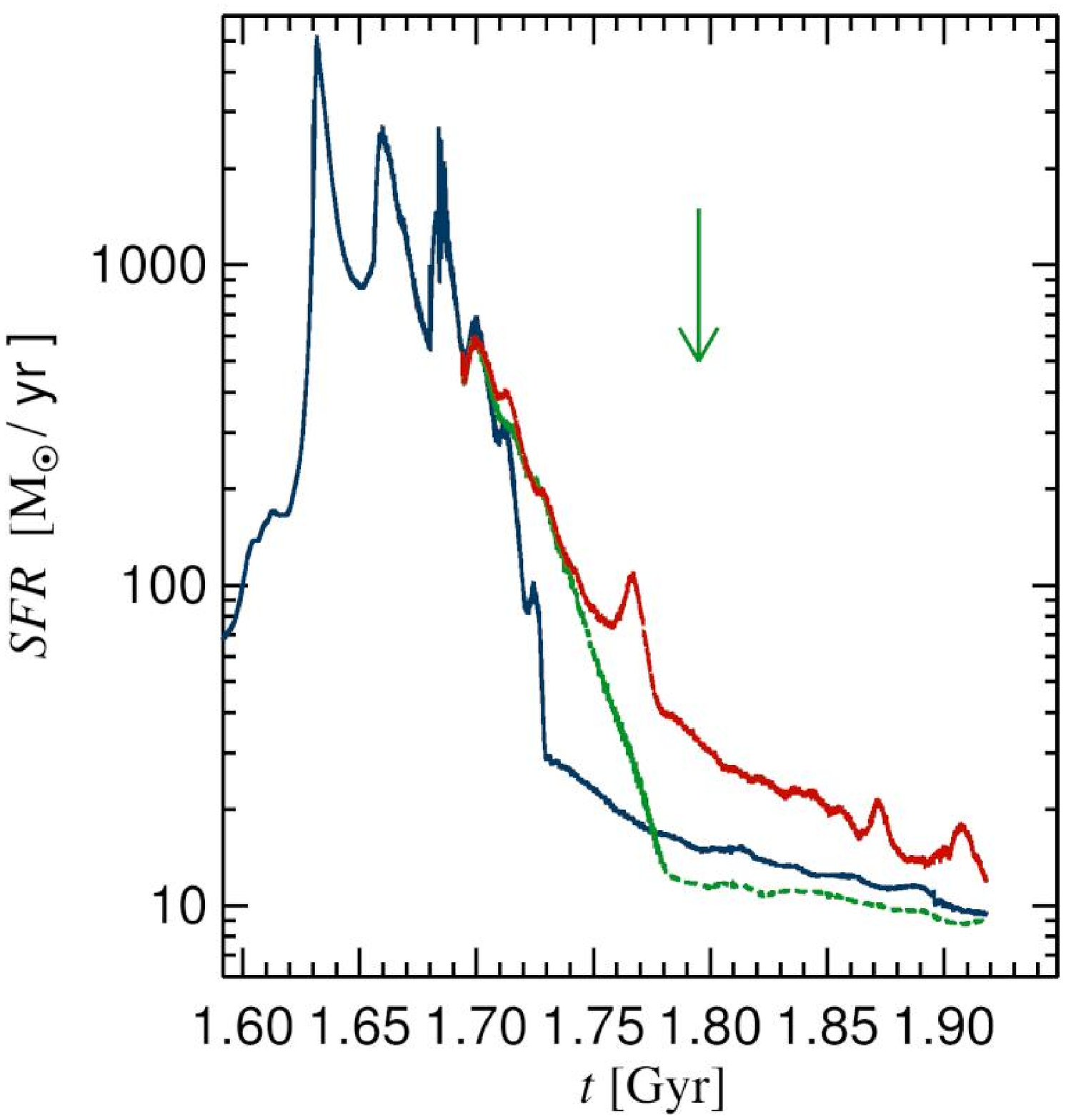}}}
\caption{BH mass (left-hand panel), bolometric luminosity (central
  panel) and total SFR (right-hand panel) as a function of time during
  a major merger of two gas-rich galaxies. The blue continuous lines
  are for the run where after the BH coalescence no recoil is imparted
  on the remnant BH. Green dashed and red triple dot-dashed lines are
  for the simulations where the remnant BH is kicked with $v_{\rm
    kick} \sim 3253\,{\rm km\,s^{-1}}\sim 0.9\,v_{\rm esc}$ and
  $v_{\rm kick} \sim 3535\,{\rm km\,s^{-1}} \sim v_{\rm esc}$,
  respectively. The green arrows indicate when the gravitationally
  recoiled BH returns to the centre in the case of lower kick
  velocity.}
\label{mbh_merger}
\end{figure*}

In Figure~\ref{mbh_merger}, we show the time evolution of BH mass, bolometric
luminosity and SFR of the merging system. The selected time sequence starts
before the two BHs merge, which happens at $\sim 1.69\,$Gyrs. In each panel,
the blue continuous lines denote the case where the remnant BH does not
experience a gravitational recoil, while the green dashed lines and the red
triple dot-dashed lines are for the simulations where the remnant BH is kicked
with $\sim 0.9\,v_{\rm esc}$ and $\sim v_{\rm esc}$, respectively. A number of
interesting features can be seen in this figure: {\it i)} Prior to
coalescence, both BHs grow rapidly, reaching the Eddington limit; {\it ii)}
During this period, the merging system also enters into a starburst phase,
with a SFR peaking at $4000\, {\rm M}_\odot\,{\rm yr}^{-1}$; {\it iii)} Once
the remnant BH is gravitationally recoiled its accretion rate drops,
especially in the case where the BH leaves the dense central regions (i.e.~for
$v_{\rm kick} \sim v_{\rm esc}$), implying that the AGN will have a much lower
bolometric luminosity and will grow less in mass; {\it iv)} Because the
remnant BH is wandering away from the centre, it becomes much less efficient
in regulating the central properties of the host galaxy. Thus, in the case of
recoiled BHs, the central starburst activity will be prolonged, with more
young stars formed in situ; {\it v)} The extended star formation activity has
a direct impact on the BH growth once it returns to the centre: more stars
have already formed, leaving less gas to fuel further BH accretion. This
explains why the BH mass in the case of the $\sim 0.9\,v_{\rm esc}$ kick stays
much lower and cannot catch-up with that of the stationary BH, even though it
returns to the centre already after $10^8\,$yrs; {\it vi)} While the amount of
stars formed has a large impact on the BH growth, it does not significantly
affect the total stellar mass, which changes by a few percent at most. We
hence find that the gravitational recoil for a single merger could in
principle introduce more scatter in the BH mass (by up to factor of $\sim 2$)
than in the bulge mass when considering BH mass - bulge mass scaling
relation (see \citet{Blecha2011} for a comprehensive study of the scatter
in the BH mass - host galaxy scaling relations for different simulated merging
systems and e.g. \citet{Volonteri2007} who reaches simular conclusions
adopting semi-analytical techniques).

While the findings discussed above should reflect the general characteristics
of gas-rich mergers of galaxies, we would however like to stress that the
quantitative details will be very sensitive to the physics of BH binary
hardening, as well as to star formation and feedback. In our simulations, we
cannot follow the BH binary hardening due to insufficient spatial
resolution. Instead, we simply assume that the BH coalescence happens rapidly
in a gas-rich environment. This is a plausible assumption, but obviously not
guaranteed to be the case. If the final BH binary hardening should take
longer, this would have a significant impact on the results. From the
left-hand panel of Figure~\ref{mbh_merger}, we can infer that during a very
short time interval of the order of $\le 5 \times 10^7\,$yrs, from the merger
of the galactic cores to the moment where the BH growth becomes
self-regulated, the BH grows rapidly. Thus, a delay in the BH coalescence
relative to what we have assumed here, and hence a delay in the moment the
recoil occurs, can significantly reduce the mass difference between a
stationary and a recoiled BH. Similarly, the efficiency of star formation in
the innermost regions will affect the amount of gas that is still available
for accretion once the BH returns to the centre.

\section{Discussion and Conclusions} \label{Conclusions}

In this study, we have used numerical simulations to discuss the complex
interplay between the baryonic component of gas-rich galaxies and the dynamics
of supermassive BHs recoiling due to a gravitational wave induced binary
merger. Our analysis has focused on understanding how BH accretion and
feedback processes can possibly modify the orbit and return timescale of
recoiled BHs. This question has important implications both for the assembly
history of the population of supermassive BHs as well as possible detections
of displaced AGN in galaxies.

In our simulations, we have deliberately chosen massive and gas-rich systems,
which could be representative of high redshift progenitors of present day
ellipticals and brightest cluster galaxies. Using isolated disc galaxies, we
were able to study the orbital evolution of kicked BHs for a variety of
different assumptions about BH accretion and feedback, and about the treatment
of the interstellar gas. We have then extended the analysis with simulation
models of major galaxy mergers, yielding a more realistic accounting of the
perturbed state of the systems that are expected to host merger recoil events
of BHs. Our main conclusions from these simulations are as follows:

\begin{itemize}

\item At the centre of galaxies, the gravitational potential is strongly
  dominated by the baryonic component. The expected escape velocity will thus
  generally be significantly larger than that from the dark matter alone. It
  will depend on the detailed assumptions for the spatial distribution and
  dynamical evolution of the baryonic component at the host galaxy centre,
  which in turn can be affected by BH feedback as well as dynamical heating by
  the motion of the BH. For the compact remnants of galaxy mergers the escape
  velocity will be larger by a factor of a few, and the kick velocities
  required for a significant displacement of the supermassive BHs will need to
  be comparable to the central escape velocity, due to the steepness of the
  potential.

\item Recent claims that massive high-redshift galaxy spheroids and discs are
  significantly more compact than their low-redshift counterparts should thus
  have a large effect on the expected trajectory of recoiling BH merger
  remnants. For all but the most extreme and rather unlikely values of the
  kick velocities, supermassive BHs should thus not be removed from the rather
  massive galaxies which appear to host the bulk of the supermassive BH
  population and the displacements should be mostly moderate (less than a few
  kpc) and rather short-lived.

\item During mergers of gas-rich systems, where it is very likely that the BHs
  will grow rapidly due to large amounts of gas being funnelled towards the
  centre, gravitational recoils increase the scatter in BH mass - host galaxy
  scaling relationships predicted by the simulations. In particular, the BH
  mass is very sensitive  to the occurrence of recoil, and can be lower by up
  to a factor of a few in respect to the mass of a BH which does not
  experience any kick. We note however that the exact amount of BH mass
  reduction is very dependent on the BH binary hardening timescale and also on
  the efficiency of the starburst to consume the central gas supplies.

\item Unfortunately, the strong sensitivity of the dynamics of a recoiling BH
  on the detailed spatial distribution and thermal state of the baryonic
  component, which in turn depends on the details of the feedback of the
  accreting BHs, adds another level of complexity to predictions of the
  expected distribution of the luminosity and the displacement of off-centred
  AGN.

\item The overall amount of accretion and therefore also the luminosity of a
  recoiling BH depends sensitively on the distribution and thermal state of
  gas down to scales smaller than can be resolved by our simulations. As in
  previous work, we have thus parametrised the accretion rate as
  Bondi-Hoyle-Lyttleton accretion with a parameter $\alpha$ to take into
  account that the simulations cannot fully resolve the multi-phase medium of
  the interstellar gas. If the recoiling BH leaves the dense multi-phase gas
  $\alpha =100$ probably overestimates the accretion rate somewhat. We have
  thus also tested what happens if we assume $\alpha =1$,
  i.e.~Bondi-Hoyle-Lyttleton accretion as derived from the actual density and
  temperature distribution of the gas in the simulation. This should now
  underestimate the actual accretion rate. The corresponding reduction of the
  accretion rate and luminosity is about a factor five -- much weaker than
  linear because of the self-regulating effect of the feedback on the
  accretion rate. In reality, the accretion rates and the corresponding
  luminosities should lie somewhere in between. The generally rather large accretion
  rates of BHs if recoiled within gas-rich discs predicted by our simulations
  would bode well for the possibility to detect off-centred AGN, and may give
  added confidence in the recent claims of such detections.

\end{itemize}

\section*{Acknowledgements} 
We would like to thank Jim Pringle for very useful discussions and comments on
the manuscript and Giuseppe Lodato for enthusiastic discussions on the
topic. DS acknowledges a Postdoctoral Fellowship from the UK Science and
Technology Funding Council (STFC) and NASA Hubble Fellowship through grant
HST-HF-51282.01-A. MH was partially supported by STFC grant LGAG 092/
RG43335. Simulations were performed on the Cambridge High Performance
Computing Cluster DARWIN in Cambridge (http://www.hpc.cam.ac.uk).

\bibliographystyle{mn2e}

\bibliography{paper}

\appendix

\section{Numerical issues} \label{appen}

If the gravitationally recoiled BH leaves the dense multi-phase medium while
on its orbit through the galaxy, the accretion rate might be overpredicted by
equation~(\ref{Bondi_eq}) when $\alpha=100$ is used \citep[see
  also][]{Booth2009}. A lower gas accretion rate and hence a lower associated
BH feedback may then possibly change the impact of the recoiled BH on its
surroundings, which in turn then leads to a different BH trajectory.

To explore these issues and test the robustness of our results, we have
performed three additional simulation studies. In one run we reduce the $\alpha$
parameter to $1$, thereby providing a lower limit to the expected accretion.
In the second we increase the numerical resolution by a factor of $8$ in
particle number for each galactic component, resulting in twice as high spatial
resolution per dimension, equal to $30\,h^{-1} \, {\rm pc}$ for the bulge and
the disc, and to $1\,h^{-1} \, {\rm kpc}$ for the dark matter halo. In the
third run we choose a smaller initial BH mass of $5 \times 10^7\,h^{-1}{\rm
  M}_\odot\,$ (corresponding to $\sim 10^{-3} M_{\rm bulge}$) and the kick
velocity of $500\,{\rm km\,s^{-1}}$.  

\begin{figure}\centerline{
\includegraphics[width=7.5truecm,height=7truecm]{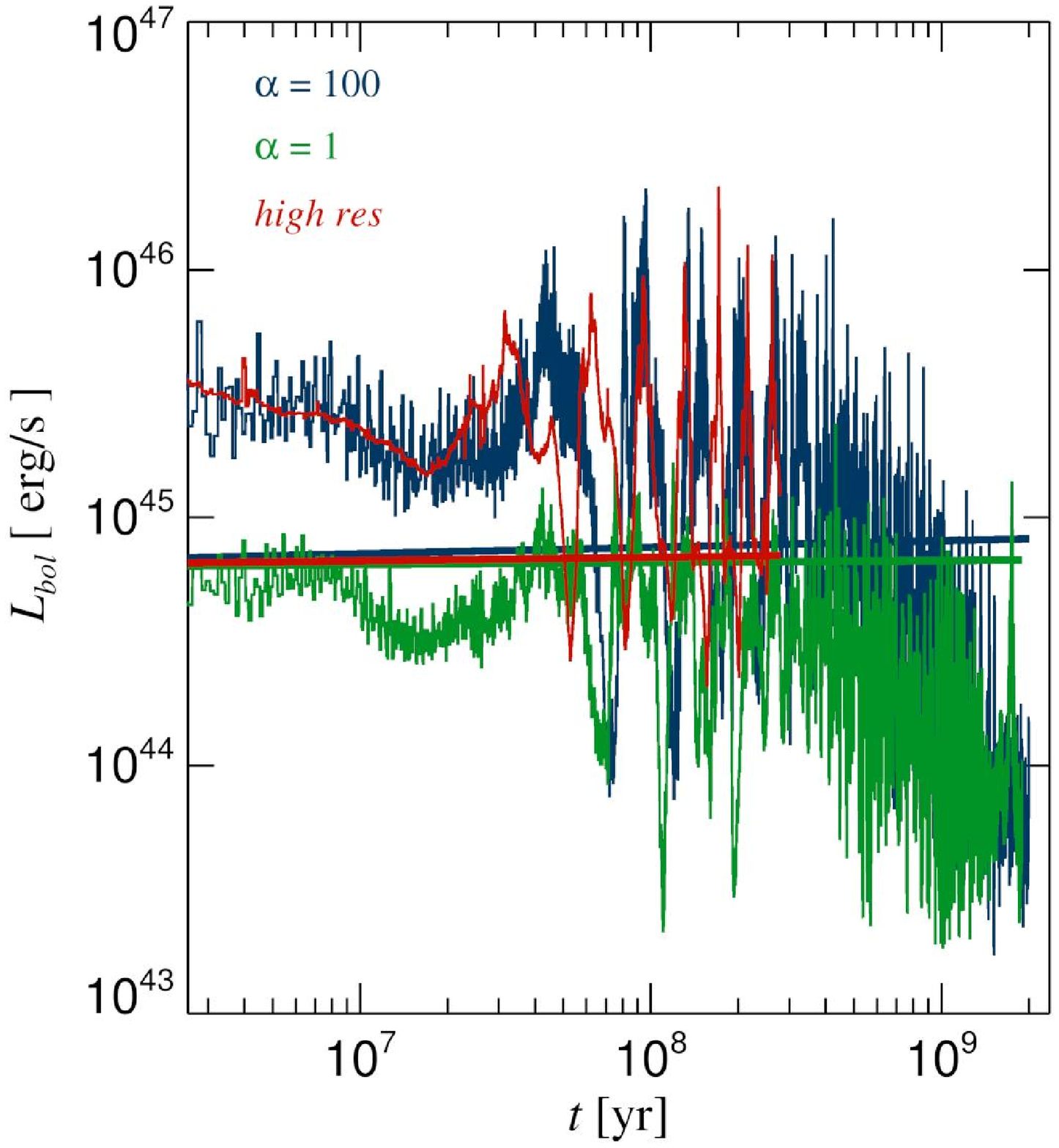}}
\caption{Bolometric luminosity as a function of time
  for simulations with two different values for the $\alpha$ parameter in
  equation~(\ref{Bondi_eq}): $\alpha  = 100$ (blue line) and $\alpha = 1$
  (green line). The same galaxy has also been simulated with twice as high
  spatial resolution and $\alpha = 100$ (red line). Straight lines denote an
  accretion rate equal to $0.01$ of the Eddington rate. } 
\label{Lbol_alpha}
\end{figure}

In Figure~\ref{Lbol_alpha} we show the bolometric luminosity of a BH which has
been kicked in the plane of the galaxy with $700\, {\rm km\, s^{-1}}$. The
blue line is for our default simulation, as shown in the central panel of
Figure~\ref{mbh_iso}. The green line is for a test run performed with $\alpha
=1$, and the red line is the result of our higher resolution simulation. For
the case of $\alpha = 1$ the accretion rate and thus the bolometric luminosity
is lower than for the simulation with $\alpha = 100$, with the bolometric
luminosity being reduced by a factor of $\sim 5$ on average. We note, however,
that $\alpha = 1$ provides clearly a lower limit of the expected
Bondi-Hoyle-Lyttleton rate, given that the recoiled BH trajectory is confined
mostly within the dense gaseous disc, which has a multi-phase structure. In
either case we expect that the recoiled BH can still have several radiatively
efficient episodes while it travels through the gas-rich disc, but these will
occur during shorter time intervals of a few $10^7\,$yrs. From
Figure~\ref{Lbol_alpha} it can also be seen that the bolometric luminosity for
our lower resolution run is in excellent agreement with the finding of the
higher resolution simulation. This indicates that our subresolution model for
BH growth and feedback gives numerically robust and convergent results.
    
\begin{figure}\centerline{
\includegraphics[width=7.5truecm,height=7truecm]{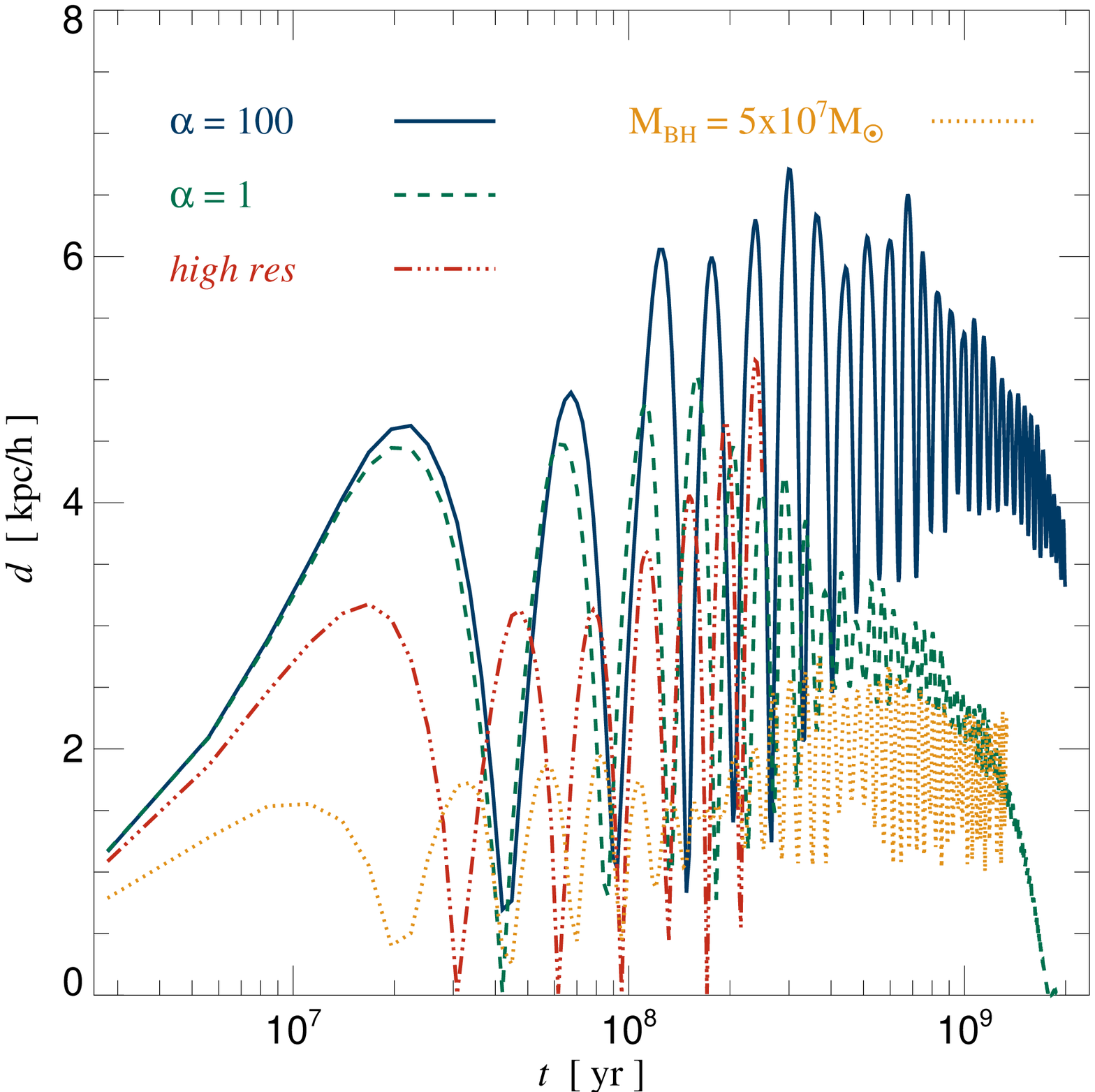}
}
\caption{BH distance from the minimum of the potential as a function of time
  for the simulations where $\alpha = 100$ in equation~(\ref{Bondi_eq})
  (blue continuous line) and where $\alpha\,=\,1$ (green dashed 
  line). The same galaxy has also been simulated with twice as high
  spatial resolution and $\alpha = 100$ (red triple dot dashed
  line). The time evolution of the BH distance for a simulation where the initial
    BH mass is set to $5 \times 10^7\,h^{-1}{\rm M}_\odot\,$ and the kick
    velocity is $500\,{\rm km\,s^{-1}}$ is shown as well (orange dotted line),
  confirming that BH feedback delays the return timescale also for less
  massive BHs.}
\label{bhdist_alpha}
\end{figure}

Figure~\ref{bhdist_alpha} shows the BH distance from the minimum of the
potential for the same set of simulations as in Figure~\ref{Lbol_alpha},
  and additionally for a simulation with a smaller initial BH
  mass. At first the BH orbits for $\alpha =1$ and $\alpha =100$ are very
similar, but after $10^8\,$yrs the recoiled BH in the case of $\alpha=1$
reaches smaller apocentric distances. Nonetheless, we observe that also in the
case of $\alpha = 1$ a plume of hot, low density gas develops in the wake of
the BH, which forces the BH into a retrograde orbit with respect to the
galactic disc. Given that the BH accretion rate is lower, the magnitude of
this effect is somewhat diminished and thus the BH returns to the centre on a
shorter timescale, which is however still rather long $\sim 1.6\times
10^9\,$yrs. Note that this findings means that our qualitative findings are
very robust regardless of the exact value of the $\alpha$ parameter, which is
very encouraging. We furthermore confirm that also for a BH with an
  initial mass ten times smaller, BH feedback causes the BH to follow a retrograde
  orbit and thus results in an analogous delay in the BH return timescale.  

The BH orbit for the higher resolution simulation is more different than
perhaps may have been expected, given that the BH mass growth is very
similar. The reason for this can be attributed to the detailed shape of the
gravitational potential, which is resolved down to smaller scales in the
higher resolution run. As a result, the central potential is about $10\%$
deeper than in the lower resolution simulation. This difference is sufficient
for the recoiled BH to reach only $\sim 3.2 \,h^{-1} \, {\rm kpc}$ instead of
$\sim 4.6 \,h^{-1} \, {\rm kpc}$ at the first apocentre.

\end{document}